\documentclass[%
reprint,
 amsmath,amssymb,
 aps,
]{revtex4-1}

\usepackage{graphicx}
\graphicspath{{./pics/}}
\usepackage{dcolumn}
\usepackage{bm}

\usepackage{upgreek}
\usepackage{placeins} 

\begin{document}


\title{Local phase transitions in a model of multiplex networks with heterogeneous degrees and inter-layer coupling}

\author{Nedim Bayrakdar}
\email{bayrakdar.nedim@gmail.com}
\affiliation{Instituut-Lorentz for Theoretical Physics, Leiden Institute of Physics, University of Leiden, Niels Bohrweg 2, 2333 CA Leiden, The Netherlands}

\author{Valerio Gemmetto}
\email{gemmetto@lorentz.leidenuniv.nl}
\affiliation{Instituut-Lorentz for Theoretical Physics, Leiden Institute of Physics, University of Leiden, Niels Bohrweg 2, 2333 CA Leiden, The Netherlands}

\author{Diego Garlaschelli}
\email{garlaschelli@lorentz.leidenuniv.nl}
\affiliation{Instituut-Lorentz for Theoretical Physics, Leiden Institute of Physics, University of Leiden, Niels Bohrweg 2, 2333 CA Leiden, The Netherlands}
\affiliation{IMT School of Advanced Studies, Piazza S. Francesco 19, 55100 Lucca, Italy}

\date{February 22, 2018}

\vspace{2in}
\begin{abstract}

\noindent 
Multilayer networks represent multiple types of connections between the same set of nodes. Clearly, a multilayer description of a system adds value only if the multiplex does not merely consist of independent layers, i.e. if the inter-layer overlap is nontrivial. On real-world multiplexes, it is expected that the observed overlap may partly result from spurious correlations arising from the heterogeneity of nodes and partly from true interdependencies. However, no rigorous way to disentangle these two effects has been developed. In this paper we introduce an unbiased maximum-entropy model of multiplexes with controllable node degrees and controllable overlap. The model can be mapped to a generalized Ising model where the combination of node heterogeneity and inter-layer coupling leads to the possibility of local phase transitions. In particular, we find that an increased heterogeneity in the network results in different critical points for different pairs of nodes, which in turn leads to local phase transitions that may ultimately increase the overlap. The model allows us to quantify how the overlap can be increased by either increasing the heterogeneity of the network (spurious correlation) or the strength of the inter-layer coupling (true correlation), thereby disentangling the two effects. As an application, we show that the empirical overlap in the International Trade Multiplex is not merely a spurious result of the correlation between node degrees across different layers, but requires a non-zero inter-layer coupling in its modeling.

\bigbreak


\end{abstract}

\pacs{Valid PACS appear here}
\maketitle

\section{Introduction}
The wide variety of different phenomena that occur around us are often the result of systems that emerge and (self-)organize dynamically. These systems consist of a multitude of basic constituents interacting with each other via complicated patterns. This abstract notion of a network allows its application to a wide array of systems. Examples of such systems include social networks, transportation networks, biological networks, financial networks, and technological networks. In social networks, for example, the individuals (or larger organizations) of the population can be represented by nodes and the social ties or relations among individuals (or larger organizations) can be represented by links between these nodes, sexual relations among adults, or simply the belonging to common institutions\cite{krackhardt1987cognitive,padgett1993robust}. The study of these networks may increase our understanding of a variety of processes, such as the spreading of sexually transmitted diseases or the diffusion of knowledge. 

A straightforward and classical approach is to map each constituent within a system onto a single node, and to map each interaction between pairs of constituents of the system onto a link of a single type regardless of the interaction's nature. All of the network's links in this approach are treated on equal footing, making it a single-layer network representation, and that may be a oversimplification that fails to capture the details of a multi-relational system. The inability to properly represent multi-relational systems using single-layer networks has lead to an increasing amount of attempts in developing a framework to study these systems using \textit{multilayer networks} \cite{de2013mathematical,battiston2014structural,kivela2014multilayer,battiston2017new,boccaletti2014structure}. Multilayer networks allow us to describe these multi-relational systems by representing each type of relationship by a layer, where a constituent of the system (node) may have different types of relationships to other constituents. Returning to the example of social networks, the different type of relationships between people such as sexual relations, friendship, coworker-ship, et cetera would each be represented by links in different layers\cite{verbrugge1979multiplexity}. 

In recent years, there has been increased attention to models that are probabilistic in nature and are also known as random graph models.  
There are a wide variety of random graph models \cite{erdos1960evolution,albert2002statistical,barabasi1999emergence,watts1998collective} and one such model is the  exponential random graph model (ERGM)\cite{holland1981exponential,besag1974spatial,frank1986markov,contractor2006testing,wasserman1994social,carrington2005models,park2004statistical}. Exponential random graph models are used commonly within the social network analysis community and have been around for a long time. It's a very general theory that provides great predictive power. However, J. Park and M. E. Newman showed that the ERGM can be derived using maximum entropy principles and showed that the model is equivalent to applying the principles of statistical mechanics to networks. This allows us to utilize techniques that are common in statistical physics. In the ERGM we choose the probability distribution over graphs such that this probability distribution maximizes the entropy. This maximization is performed while the expected values of certain chosen graph properties are constrained to be equal to desired values. 

In studies where real-world multilayer networks have been modeled using the ERGM, the different layers of the networks are often assumed to be independent of each other \cite{gemmetto2015multiplexity,gemmetto2016multiplexity}. The main reason for this assumption is that any introduction of interdependence between the layers into the ERGM will greatly sophisticate the mathematics. There are many properties of a multilayer network that encode this interdependence. Two such properties are the \textit{overlap} and the \textit{multiplexity} \cite{gemmetto2015multiplexity,battiston2014structural}. The overlap and the multiplexity essentially contain the same information and capture the correlation of the node connectivity patterns in two or more layers.

For example, in some social network people communicate with their friends through multiple means of communications, such as talking on the phone, sending emails or sending instant text messages. In this example, the layer that represents communication through email has a significant overlap with the layer of communication through text messages. A more specific example is the study of the World Trade Network using the ERGM on a multilayer network with the assumption that the layers are independent, which showed that the observed overlap is significantly different from the overlap predicted by the model \cite{gemmetto2015multiplexity}. This result is not unexpected, since one can imagine that the trade of a certain product between two countries may increase/decrease the possibility of the trade of a \textit{different} product between the same two countries. Other examples of networks displaying a significant overlap are airport networks, on-line social games, collaboration networks and citation networks \cite{szell2010multirelational,cardillo2013emergence,menichetti2014weighted}. 

A large part of the observed overlap in many of these networks could actually be created entirely by chance instead of resulting from interlayer dependence \cite{gemmetto2015multiplexity,gemmetto2016multiplexity}, there will be an increased probability of a link between two nodes being present in multiple layers while the probability of a link occurring in one layer does not necessarily influence the presence of a link occurring in another layer. The measured overlap of the network therefore consists of a part resulting from 'spurious' coupling between the layers and of a part resulting from genuine coupling between the layers. This effect of spurious coupling increases as the density and/or heterogeneity of the degrees of the network increases. Real world networks are often dense and have strongly heterogeneous degrees and therefore the assessment of interlayer coupling in these real world networks are severely affected.

The focus of this paper is the introduction of interdependencies between the layers of a multilayer network in the ERGM through the explicit inclusion of the overlap. This inclusion of the overlap into the ERGM might  aid us in understanding what (higher order) properties of network structure may be (highly) dependent on the overlap. Additionally, it will aid us with distinguishing between the overlap in the network due to the correlation of single-node properties across layers and the overlap due to a genuine coupling between the layers. Finally, it will allow us to generate null models with the desired amount of spurious overlap and genuine overlap. It turns out that this problem is mathematically very similar to solving the Ising model on a complete graph (which is also known as the Curie-Weiss model).

The rest of the paper is organized as follows: in Section \ref{sec: Background Theory} we mathematically define quantities and models that are relevant to this paper. This includes the derivation of a model where the layers of the multiplex network are independent and our original model where the layers of the multiplex are interdependent through the inclusion of the overlap. Section \ref{sec: Phase transitions} contains a discussion regarding the possible phase transitions of our model. In Section \ref{sec: Numerical Analysis} we explore our model by using various numerical methods. In Section \ref{sec: Empirical Analysis} we briefly analyze the International Trade Network and show that the empirical overlap in this real world network is not merely the result of the heterogeneity of the network but requires a nonzero coupling between the layers in its modeling.

\section{Background Theory}
\label{sec: Background Theory}
This section contains the formal definitions and descriptions of a few selected network theoretical notions and models that are relevant to this paper.

\subsection{Single-layer network definitions}
\label{subsec: Single Layer Network Definitions}
The relations between the constituents of a system can generally be classified as either \textit{binary and undirected}, \textit{binary and directed}, \textit{weighted and undirected} or \textit{weighted and directed}. 
In a system where the relations between the constituents are directed, the links in the network representation have a direction associated with them while the links in a undirected network have no orientation. In a system where the relations are weighted, the links in the network representation have a number (the weight) assigned to each link while the links in a binary network simply exist or do not exist which can be signified by a binary number assigned to the links. We will limit our further discussion to the \textit{binary and undirected} case. 

A binary undirected network can be defined as a graph which is an ordered pair $G = (V, E)$, where $V = \{v_1,v_2,...,v_N\}$ is a set of $N$ \textit{vertices} or \textit{nodes}, and $E$ is a set of \textit{unordered} pairs of different vertices called \textit{edges} or \textit{links}. Note that the definition of $E$ depends on the relevant class of relations between the constituents of the system. The vertex $v_i\in V$ may be referred to as simply \textit{i} throughout the rest of the paper.  If $(i,j) \in E$, the vertices $i$ and $j$ are said to be connected, and may be referred to as \textit{neighbors} of each other. The number of links $L$ of the graph is given by the cardinality of $E$: $L = |E|$.

\paragraph*{Matrix Representation} A graph $G$ is represented by its \textit{adjacency matrix} $G = \{g_{ij}\} $. This is an $N \times N$ matrix where 
\begin{equation}
g_{ij} =
	\begin{cases}
	1, & \text{if}\ (i,j) \in E \\
	0, & \text{otherwise}
	\end{cases}
\end{equation}
We defined $E$ to contain pairs of \textit{different} vertices, which means that a vertex can not have a connection to itself. It is natural to then define the diagonal elements as $g_{ii} \equiv 0$.
Since we limit our discussion to undirected graphs, the adjacency matrix is always symmetric: $g_{ij} = g_{ji}$ and it therefore contains $N(N-1)/2$ elements that fully specify the matrix and ultimately the graph.

\paragraph*{Degrees and degree distribution}
One of the main topics in the analysis of complex networks is the identification of nodes that play a central role in the structure of the network \cite{freeman1977set} . There are a variety of measures that characterize the structural importance of a node in a network. The degree $k_i$ is defined as the number of connections node $i$ has to other nodes in the network.
\begin{equation}
k_i  = \sum_{j\neq i}^N g_{ij}
\end{equation}
The set $\{ k_i\}$ of degrees is called the \textit{degree sequence} of the network.
The degree distribution $P(k)$ is defined to be the fraction of nodes in the network with degree $k$.
\begin{equation}
P(k) = \frac{\textrm{\# of nodes whose degree equals } k}{N}
\end{equation}
Real-world networks consistently show a degree distribution with heavy tails, where the degrees vary over a broad range, often spanning several orders of magnitude\cite{clauset2009power,barabasi1999emergence}. A large number of vertices often have a small number of links to other vertices while a small minority of the vertices have a large number of links to other vertices.
An example of a heavy tail distribution is the power-law distribution, which we will later use to study our model. 

\subsection{Multiplex network definitions}
\label{subsec: Multiplex network definitions}
A binary undirected multiplex network can be defined in terms of the previously defined single layer networks. A multiplex network is a set $\vec{G} = \{G_\alpha\}^M_{\alpha=1}$ of $M$ single layer undirected binary graphs $G_{\alpha} = (V, E_\alpha)$ which share the same set of $N$ nodes. In the context of multilayer networks, $G_{\alpha}$ is called a layer of $\mathcal{M}$ and may be referred to as simply $\alpha$ throughout the rest of the paper. A multiplex network is a specific type of multilayer network which does not allow interlayer connections between two layers $\alpha$ and $\beta$ where $\alpha \neq \beta$.

\paragraph*{Matrix Representation}
The layer $G_{\alpha}$ and its intra-layer links can then be represented by its adjacency matrix $G_{\alpha} = \{ g_{ij}^\alpha \}$. This is an $N \times N$ matrix where 

\begin{equation}
g_{ij}^\alpha =
	\begin{cases}
	1, & \text{if}\ (i^\alpha,j^\alpha) \in E_\alpha \\
	0, & \text{otherwise}
	\end{cases}
\end{equation}

\paragraph*{Multilinks in Multiplex Networks}
\label{par: Multilinks}
In order to capture the information regarding the presence of the links between the pair of nodes $(i,j)$ in any of the $M$ layers, we define the object
\begin{equation}
\label{eq: multilink definition} m_{ij} \equiv \{g_{ij}^1,g_{ij}^2,\ldots, g_{ij}^M\}
\end{equation}
which is also known as the \textit{multilink} of $(i,j)$. Additionally, we define the set $\mathcal{M}_{ij}$ as the set that contains all possible configurations of $m_{ij}$ which therefore contains $2^{M(M-1)/2}$ elements.

\paragraph*{Degrees}
In Subsection \ref{subsec: Single Layer Network Definitions} we mentioned several characterizations that measure the structural importance of a node in a single-layer network. To study the structural importance of specific nodes in a multilayer network, one can extend these single-layer characterizations to multilayer networks. Since we limited our discussion in Section \ref{subsec: Single Layer Network Definitions} to the degree, we shall solely discuss multilayer measures of structural importance that are extensions of the single-layer degree.

The degree of a node $i \in V$ of a multiplex network $\vec{G}$ is the object

\begin{equation}
\label{eq: Multiplex Degree}
\vec{k}_i = (k_i^1,k_i^2,...,k_i^M)
\end{equation}
where 
\begin{equation}
k_i^\alpha = \sum_{j\neq i}^Ng_{ij}^\alpha 
\end{equation}
is the degree of the node $i$ in the layer $\alpha$ \cite{berlingerio2011foundations,battiston2014structural} . The vector definition of the degree of a node makes it difficult to compare the structural relevance of each node. A scalar measure would be more suitable for making such comparisons. One such scalar measure is the \textit{layer-average degree}:
\begin{equation}
\label{eq: layer average degree}
\overline{k}_i = \frac{1}{M}\sum_{\alpha=1}^M k_i^\alpha 
\end{equation}
which is the degree of node $i$ averaged over the $M$ layers.

\paragraph*{Overlap}
\label{subsec: Overlap}
We mentioned that there are many properties that encode the interdependence between the layers of a multilayer network, and that we will limit our discussion to one such property: the overlap. The overlap $O^{\alpha \beta}$ between two layers $\alpha$ and $\beta$ is defined as the number of links that appear in both layer $\alpha$ and $\beta$ \cite{bianconi2013statistical,szell2010multirelational} :
\begin{equation}
\label{eq: pair overlap}
O^{\alpha\beta} = \sum_{i<j}g_{ij}^\alpha g_{ij}^\beta.
\end{equation}
The \textit{global overlap} $O$ is defined as the sum of $O^{\alpha\beta}$ over all pairs of layers:
\begin{equation}
\label{eq: global overlap}
O = \sum_{\alpha < \beta }\sum_{i<j} g_{ij}^\alpha g_{ij}^\beta
\end{equation}
As the names of these properties suggest, they are a measure of how overlapping the layers of the multiplex network are. The individual terms in \eqref{eq: pair overlap} are only nonzero when a link exists between two nodes $i$ and $j$ in both layer $\alpha$ and $\beta$.

\subsection{Exponential Random Graph Model}
\label{subsec: ERGM}
The exponential random graph model (ERGM) is an ensemble model, which means that the model is not defined as a single (multiplex) network, but a probability distribution over many possible (multiplex) networks. Given the observed (or desired) values $z_i^*$ of a collection of graph observables $\{z_i\}$ of some real-world network where $i=1,\ldots, K$, the exponential graph model generates a probability distribution over graphs that assigns a higher probability to graphs that have values for the graph observables that are similar to the ones of the real-world network. This method provides us with overall framework for modeling, and allows us to incorporate mechanisms that might be responsible for features observed in empirical studies of networks.
\bigbreak
Let $\mathcal{G}^M_N$ be the set of (binary undirected) multiplex networks consisting of $N$ vertices and $M$ layers (note that this set includes single-layer networks for $M = 1$), let $\vec{G}=\{G_1,G_2,...,G_M\} \in \mathcal{G}^M_N$ be a multiplex network in that set of multiplex networks, and let $P(\vec{G})$ be the probability of $\vec{G}$ within the ensemble. One would ideally choose $P(\vec{G})$ such that the expectation value of each graph observable $z_i(\vec{G})$ is equal to the observed value. This type of probability distribution is also referred to as a \textit{canonical ensemble}. The ideal probability distribution is the one which maximizes the Gibbs entropy
\begin{equation}
S = -\sum_{\vec{G}\in \mathcal{G}^M_N} P(\vec{G})\ln{P(\vec{G})}
\end{equation}
The maximization of the entropy is moreover constrained by a number of statistical observables $z_i(G)$ where $i = 1,\ldots,K$ for which one assumes one has the observed (or desired) values $z_i ^*$ 
\begin{equation}
\label{eq: ERGM constraint}
 z_i^* = \langle z_i \rangle 
\end{equation}
where 
\begin{equation}
\langle z_i \rangle = \sum_{\vec{G}\in \mathcal{G}^M_N}P(\vec{G})z_i(\vec{G})
\end{equation}
and by the normalization condition 
\begin{equation}
\sum_{\vec{G}\in \mathcal{G}^M_N}P(\vec{G}) = 1.
\end{equation}
The maximization of the entropy function is done by introducing a Lagrange multiplier $\theta_i$ for every constraint $\langle z_i \rangle = z_i^* $ and $\alpha$ for the normalization condition. This leads to the solution
\begin{equation}
\label{eq: ERGM Probability}
P(\vec{G},\vec{\theta}) = \frac{e^{-H(\vec{G}, \vec{\theta})}}{Z(\vec{\theta})}
\end{equation}
where $H(\vec{G}, \vec{\theta})$ is the graph Hamiltonian 
\begin{equation}
\label{eq: ERGM Hamiltonian}
H(\vec{G}, \vec{\theta}) \equiv \sum_{i}\theta_i z_i(\vec{G}) = \vec{\theta} \cdot \vec{z}(\vec{G})
\end{equation}
and $Z$ is the partition function whose form is imposed by the normalization condition
\begin{equation}
\label{eq: ERGM Partition Function}
Z(\vec{\theta}) \equiv e^{\alpha+1} = \sum_{\vec{G} \in \mathcal{G}^M_N} e^{-H(\vec{G}, \vec{\theta})}.
\end{equation}
The values of the parameters $\theta_i$ that correspond to the observed (or desired) values $z_i^*$ can be found by solving the equations that are defined by the constraints \eqref{eq: ERGM constraint} :
\begin{equation}
\label{eq: Self Consistent Average}
 z_i^* = \sum_{\vec{G}\in\mathcal{G}^M_N}z_i(\vec{G})\frac{e^{-\sum_{j}\theta_j z_j(\vec{G})}}{Z(\vec{\theta})}
\end{equation}
Equations \eqref{eq: ERGM Probability}, \eqref{eq: ERGM Hamiltonian} and \eqref{eq: ERGM Partition Function} fully define the exponential random graph model. Maximizing the entropy subject to a set a set of constraints is widely used in problems with incomplete information \cite{jaynes1957information,jaynes1982rationale} . 

\subsection{Maximum-Likelihood parameter estimation}
\label{subsec: MLP}
When considering the Lagrange multipliers $\theta_i$ in the exponential random graph model as free parameters, one can study the effects that the specification of certain graph observables $z_i$ has on other aspects of network structure \cite{park2004statistical,garlaschelli2006multispecies,newman2004finding,anand2009entropy,garlaschelli2009generalized} . This approach however does not allow one to consider exponential random graph ensembles as null models of a particular real network \cite{squartini2011analytical} .  The maximum-likelihood parameter estimation can be used to generate an exponential random graph ensemble that can be used as a null model for a particular real (multiplex) network. This null model can then be used to detect statistically significant deviations of empirical structural properties of a real network from the null model. 

Suppose that we have an empirical multiplex network $\vec{G}^*$. We define the log likelihood of the multiplex $\vec{G}^*$ 
\begin{equation}
\mathcal{L}(\vec{G}^*,\vec{\theta}) = \ln{P(\vec{G}^*,\vec{\theta})} = -\ln{Z(\vec{\theta})} - \sum_{i}\theta_i z_i^*.
\end{equation}
This function has the following properties:
\begin{equation}
\label{eq: log likelihood derivative}
\frac{\partial \mathcal{L}}{\partial \theta_i} = \langle z_i \rangle - z_i^*
\end{equation}
\begin{equation}
\label{eq:MLPE covariance}
\begin{split}
\frac{\partial ^2 \mathcal{L}}{\partial \theta_i \theta_j } & = -\langle z_i z_j \rangle + \langle z_i \rangle \langle z_j \rangle. 
\end{split}
\end{equation}
Equation \eqref{eq: log likelihood derivative} means that the stationary points of $\mathcal{L}$ are precisely those $\vec{\theta} = \vec{\theta}^*$ that satisfy the constraints \eqref{eq: ERGM constraint}, such that 
\begin{equation}
\label{eq: MLPE constraint}
\langle z_i \rangle_{\vec{\theta}^*} = \sum_{\vec{G}\in\mathcal{G}^M_N}z_i(\vec{G})P(\vec{G}|\vec{\theta}^*) = z_i(\vec{G}^*) 
\end{equation}
where $\langle z_i \rangle_{\vec{\theta}^*}$ indicates that the ensemble average is evaluated at the values $\vec{\theta}^*$. Equation \eqref{eq:MLPE covariance} means that $\mathcal{L}$ is concave, since $\partial^2\mathcal{L}/\partial \theta_i \theta_j$ has the form of a negative covariance matrix and must therefore be non-positive definite \cite{coolen2017generating}. The solutions $\vec{\theta}^*$ of the coupled equations $\langle z_i \rangle_{\vec{\theta}^*} = z_i^*$ (\eqref{eq: ERGM constraint}) can therefore be found by maximizing the log likelihood $\mathcal{L}$. If $\partial^2 \mathcal{L}/ \theta_i \theta_j$ is negative definite, which will be true if the functions $z_i(\vec{G})$ are linearly independent\cite{coolen2017generating} , there will be at most one solution and it will be the unique maximum of $\mathcal{L}$. Maximizing a concave function (which is identical to minimizing a convex function) is easier than solving the system of coupled nonlinear equations in Equation \eqref{eq: Self Consistent Average}. Once the solution $\vec{\theta} = \vec{\theta}^*$ is found, it can be used to generate a null model of $\vec{G}^*$.

\subsection{Independent Layers Model}
\label{subsec: ACM}
Suppose that we measure the layer average degrees (as defined in Equation \eqref{eq: layer average degree}) of all vertices of a real multiplex $\vec{G}^*$ and we wish to create a null model of the network using the exponential random graph model in combination with the maximum-likelihood method. This model will be referred to as the \textit{average layer configuration model (ACM)}. The appropriate Hamiltonian of our exponential random graph is in this case
\begin{equation}
\label{eq: ACM Hamiltonian}
H = M\sum_{i}\theta_i \overline{k}_i = \sum_{\alpha}\sum_{i<j}(\theta_i+\theta_j)g_{ij}^\alpha 
\end{equation}
where we have multiplied the layer average degrees with $M$ for convenience. The partition function is
\begin{equation}
\begin{split}
\label{eq: Example_3 Partition Function}
Z & = \sum_{\vec{G}\in\mathcal{G}^M_N}e^{-\sum_\alpha \sum_{i<j}(\theta_i+\theta_j)g_{ij}^\alpha} 
\\ & = \sum_{\vec{G}\in\mathcal{G}^M_N}\prod_{\alpha}\prod_{i<j}e^{-(\theta_i+\theta_j)g_{ij}^\alpha} 
\\ & = \prod_{\alpha}\prod_{i<j}\sum_{g_{ij}^\alpha = 0}^1 e^{-(\theta_i+\theta_j)g_{ij}^\alpha} 
\\ & = \prod_{\alpha}\prod_{i<j}\left( 1+e^{-(\theta_i+\theta_j)}\right)
\end{split}
\end{equation}
The probability distribution over the ensemble is then given by
\begin{equation}
\label{eq: Example_3 Probability}
P(\vec{G}) = \prod_{\alpha}\prod_{i<j}\frac{e^{-(\theta_i + \theta_j)g_{ij}^\alpha}}{1+e^{-(\theta_i + \theta_j)}}.
\end{equation}
Note that the individual factors in Equation \eqref{eq: Example_3 Partition Function} do not depend on $\alpha$. The log likelihood of the multiplex $\vec{G}^*$ is 
\begin{equation}
\mathcal{L} = -M\sum_{i}\theta_i \overline{k}_i^* -\sum_{\alpha}\sum_{i<j}\ln{\left(1+e^{-(\theta_i+\theta_j)} \right)}
\end{equation}
where $\overline{k}_i^* = \overline{k}_i(\vec{G}^*)$. We want to maximize the log-likelihood for every $\theta_m$ and therefore the solution $\theta_m = \theta_m^*$ must satisfy
\begin{equation}
\begin{split}
\frac{\partial \mathcal{L}}{\partial \theta_m^*} 
 & = -M\overline{k}_m^* + \sum_{\alpha}\sum_{i\neq m}\frac{e^{-(\theta_i^*+\theta_m^*)}}{1+e^{-(\theta_i^* + \theta_m^*)}} = 0 \qquad \forall m
\end{split}
\end{equation}
or equivalently 
\begin{equation}
\label{eq: Example_3 MLP Equations}
\overline{k}^*_i = \frac{1}{M}\sum_{\alpha}\sum_{j\neq i}\frac{e^{-(\theta_i^*+\theta_j^*)}}{1+e^{-(\theta_i^* + \theta_j^*)}}  \qquad \forall i
\end{equation}
According to the maximum-likelihood principle, the empirical layer average degree $\overline{k}_i^* = \overline{k}_i(\vec{G}^*)$ of the real multiplex $\vec{G}^*$ is equal to the ensemble average $\langle \overline{k}_i \rangle_{\bm{\theta}^*}$:
\begin{equation}
\label{eq: Example_3 MLP Constraint}
\begin{split}
\overline{k}^*_i &= \langle \overline{k}_i \rangle_{\bm{\theta}^*}  
\\ &= \frac{1}{M}\sum_{\alpha}\sum_{j\neq i}\langle g_{ij}^\alpha \rangle_{\bm{\theta}^*} 
\\ &= \frac{1}{M}\sum_{\alpha}\sum_{j\neq i}p_{ij}^\alpha 
\end{split}
\end{equation}
where $p_{ij}^\alpha$ is the probability that a link occurs between node $i$ and $j$ in layer $\alpha$. From Equations \eqref{eq: Example_3 MLP Equations} and \eqref{eq: Example_3 MLP Constraint} one may infer that 
\begin{equation}
\label{eq: Example_3 Link Probability}
p_{ij}^\alpha = p_{ij} = \frac{e^{-(\theta_i^*+\theta_j^*)}}{1+e^{-(\theta_i^* + \theta_j^*)}} 
\end{equation}
The probability distribution $P(\vec{G})$ can now be written as a product over the layers:
\begin{equation}
P(\vec{G}) = \prod_{\alpha} P_{\alpha}
\end{equation}
where $P_{\alpha}$ is the probability distribution over a single layer graph ensemble:
\begin{equation}
P_\alpha = P_\alpha(G_\alpha) =  \prod_{i<j} p_{ij}^{g_{ij}^\alpha}(1-p_{ij})^{1-g_{ij}^\alpha}
\end{equation}
This means that once the maximum-likelihood Equations \eqref{eq: Example_3 MLP Equations} are (numerically) solved, each layer of the null model of $\vec{G}$ can be generated by using an \textit{average} link probability $p_{ij}$ that is equal throughout the layers. This is a consequence of exclusively constraining properties that are averages of single-layer network properties over the layers, which essentially reduces the problem to a single-layer network problem. 

\subsection{The Average Layer Plus Overlap Configuration Model}
\label{chapter: 3}
In this section we attempt to create a better model of a multiplex network that has interdependent layers by incorporating the overlap in the exponential random graph model (ERGM). This model would therefore be an improvement to the configuration model.

\subsubsection{Constructing the Hamiltonian}
\label{subsubsec: Hamiltonian}
Suppose that we measure the layer average degrees (as defined in Equation \ref{eq: layer average degree}) of all vertices and the global overlap (as defined in Equation \eqref{eq: global overlap}) of a multiplex $\vec{G}$ with $M$ layers and $N$ vertices. We wish to create a null model of the network using the ERGM in combination with the maximum-likelihood method. This model is therefore an extension of the ACM (which we discussed in the previous section) and will be referred to as the \textit{average layer plus overlap configuration model (AOCM)}. The appropriate Hamiltonian of our exponential random graph is in this case 
\begin{equation}
\label{eq: Hamiltonian}
\begin{split}
H & = M\sum_{i}\theta_i \overline{k}_i - \frac{4J}{M}\cdot O \\
& = \sum_{i<j}\sum_{\alpha=1}^M (\theta_i + \theta_j) g_{ij}^\alpha - \frac{4J}{M}\sum_{i<j}\sum_{\alpha<\beta}g_{ij}^\alpha g_{ij}^\beta 
\end{split}
\end{equation}
where $\theta_i,J$ are the Lagrange multipliers of the ERGM and $\overline{k}_i, O$ are the average layer degrees and the overlap respectively.
We have rescaled the Lagrange multipliers for later convenience. This Hamiltonian can be written as a sum over the pairs of vertices:
\begin{equation}
\label{eq: sum pair hamiltonian}
H = \sum_{i<j} h_{ij}
\end{equation}
where 
\begin{equation}
\label{eq: pair hamiltonian}
h_{ij}  \equiv \sum_{\alpha=1}^M (\theta_i + \theta_j) g_{ij}^\alpha - \frac{4J}{M} \sum_{\alpha < \beta} g_{ij}^\alpha g_{ij}^\beta 
\end{equation}
will be referred to as the \textit{pair Hamiltonian}. We will map the variables $g_{ij}^\alpha \in \{0, 1\}$ to the new variables $\sigma_{ij}^\alpha \in \{-1, 1\}$:
\begin{equation}
g_{ij}^\alpha  = \frac{1}{2}(\sigma_{ij}^\alpha +1)
\end{equation}
Applying this transformation to Equation \eqref{eq: pair hamiltonian} and taking the limit $M \rightarrow \infty$ results in the pair Hamiltonian 

\begin{equation}
\lim_{M\rightarrow \infty} h_{ij} = \sum_{\alpha = 1}^M \left( \frac{\theta_{ij}}{2} - J \right)\sigma_{ij}^\alpha - \frac{J}{M} \sum_{\alpha<\beta}\sigma_{ij}^\alpha \sigma_{ij}^\beta - \frac{JM}{2} + \frac{M\theta_{ij}}{2}
\end{equation}
where 
\begin{equation}
\theta_{ij}\equiv \theta_i + \theta_j.
\end{equation}
Note that every quantity, variable or expression will be evaluated in the limit $M\rightarrow \infty$ throughout the rest of this paper, even though it may not be stated explicitly. If we define
\begin{equation}
B_{ij} \equiv J - \frac{\theta_{ij}}{2}
\end{equation}
\begin{equation}
h_{ij}^0 \equiv -MB_{ij} + \frac{JM}{2}
\end{equation}
the pair Hamiltonian finally reduces to 
\begin{equation}
\label{eq: Curie Weiss Model}
h_{ij} = -\sum_{\alpha=1}^M B_{ij}\sigma_{ij}^\alpha - \frac{J}{M}\sum_{\alpha<\beta}\sigma_{ij}^\alpha \sigma_{ij}^\beta + h_{ij}^0.
\end{equation}
For every specific pair of nodes $(i,j)$, the variables $\sigma_{ij}^\alpha$ can be thought of as Ising spins residing on the edges of a fully connected graph with $M$ nodes, where every Ising spin interacts with every other $M-1$ spins and is coupled to a 'field' $B_{ij}$. In terms of networks, this means that for every specific pair of nodes $(i,j)$, the edges of $(i,j)$ throughout the $M$ layers are coupled to a field $B_{ij}$ and the edge of $(i,j)$ in layer $\alpha$ are coupled to the edge of $(i,j)$ in every layer $\beta$ where $\alpha \neq \beta$ with a constant coupling strength. This model is also known as the \textbf{Curie-Weiss model}, or the Ising model on a complete graph. The full Hamiltonian \eqref{eq: sum pair hamiltonian} is therefore a summation over the Hamiltonians of \textit{non-interacting} Curie-Weiss systems. 


\paragraph*{Obtaining the Partition Function}
The Hamiltonian \eqref{eq: sum pair hamiltonian} is the sum of pair Hamiltonians \eqref{eq: Curie Weiss Model}, which means that our multiplex network consists of non-interacting pairs of nodes. As a consequence, the partition function (as defined in \eqref{eq: ERGM Partition Function}) can be written as a product over the partition functions of the pairs $(i,j)$:

\begin{equation}
\begin{split}
Z = \sum_{\vec{G}\in \mathcal{G}^M_N}e^{-H} & = \sum_{\vec{G}\in \mathcal{G}^M_N} \prod_{k<l}e^{-h_{kl}} \\
& = \prod_{i<j}z_{ij}
\end{split}
\end{equation}
where $z_{ij}$ is defined as the \textit{pair partition function}, which is a sum over all $2^{M(M-1)/2}$ possible multilinks of $(i,j)$ in $\mathcal{M}_{ij}$:
\begin{equation}
\label{eq: pair partition function}
z_{ij} \equiv \sum_{m_{ij}\in \mathcal{M}_{ij}} e^{-h_{ij}}
\end{equation}
where $m_{ij}$ is the multilink of $(i,j)$ as defined in Subsection  \ref{subsec: Multiplex network definitions}, $\mathcal{M}_{ij}$ is the set of all possible configurations of $m_{ij}$. The goal is ultimately to calculate the graph probability , which can be written in terms of the pair partition function:
\begin{equation}
\begin{split}
P(\vec{G}) 
& = \prod_{i<j} p_{ij} 
\end{split}
\end{equation}
where 
\begin{equation}
p_{ij} \equiv \frac{e^{-h_{ij}}}{z_{ij}}
\end{equation}
depends on the parameters $\theta_{ij}, J$ and the multilink $m_{ij}$.
The complete partition function and graph probability can therefore be obtained by obtaining each of the (simpler) independent pair partition functions $z_{ij}$ that corresponds to its respective multilink $m_{ij}$, where each multilink can be regarded as a configuration of a Curie-Weiss system.

\subsection{The Hubbard Stratonovich transformation}
The pair Hamiltonian \eqref{eq: Curie Weiss Model} can be rewritten as 

\begin{equation}
h_{ij} = -B_{ij}\sum_{\alpha=1}^M \sigma_{ij}^\alpha - \frac{J}{2M}\left(\sum_{\alpha=1}^M \sigma_{ij}^\alpha \right)^2 + \frac{J}{2} + h_{ij}^0.
\end{equation}
By then performing Hubbard Stratonovich transformation to the pair partition function and using the Laplace theorem \cite{polya1972problems} in the limit $M\rightarrow \infty$ we obtain an explicit expression for $z_{ij}$:
\begin{equation}
\label{eq: pair partition function expression}
z_{ij} = 2^Me^{-\frac{M}{2}\theta_{ij}-2JMu_{ij}\left(u_{ij}^*-1\right)}\cosh^M{\left(2Ju_{ij}^* -\frac{\theta_{ij}}{2} \right)}
\end{equation}
where $u_{ij}^*$ is the solution to the equation
\begin{equation}
\label{eq: u_ij tanh equation}
u_{ij} = \frac{1}{2} + \frac{1}{2}\tanh{\left( 2Ju_{ij} - \frac{\theta_{ij}}{2}\right)}.
\end{equation}
A similar approach is used in \cite{park2004solution}. The full derivation of Equation \eqref{eq: pair partition function expression} can be found in Appendix \ref{A: Pair Partition Function Derivation}.
\paragraph*{Maximum likelihood}Given a particular real multiplex network $\vec{G}$ the log likelihood is 
\begin{equation}
\mathcal{L} = \ln{P(\vec{G})} = \sum_{i<j}\left(-h_{ij} - \ln{z_{ij}} \right).
\end{equation}
At a stationary point of $\mathcal{L}$, the derivatives of $\mathcal{L}$ (and therefore also $-\mathcal{L}$) with respect to every Lagrange multiplier must equal zero:
\begin{equation}
-\frac{\partial \mathcal{L}}{\partial \theta_k} = \sum_{i<j}\frac{\partial h_{ij}}{\partial \theta_k} + \sum_{i<j} \frac{\partial \ln{z_{ij}}}{\partial \theta_k} = 0,
\end{equation}
\begin{equation}
-\frac{\partial \mathcal{L}}{\partial J} = \sum_{i<j}\frac{\partial h_{ij}}{\partial J} + \sum_{i<j} \frac{\partial \ln{z_{ij}}}{\partial J} = 0.
\end{equation}
The derivatives of the log likelihood with respect to $\theta_k$ are
\begin{equation}
\label{eq: mlp equation spin variables theta}
\begin{split}
-\frac{\partial \mathcal{L}}{\partial \theta_k} 
& = \sum_{\alpha=1}^M \sum_{j\neq k}g_{jk}^\alpha - M\sum_{j\neq k}u_{jk}^* = 0
\end{split}
\end{equation}
where we've used the facts that $g_{ij}^\alpha$ and $u_{ij}$ are symmetric with respect to the indices $(i,j)$:
\begin{equation}
\sum_{i<j}g_{ij}^\alpha \delta_i^k = \sum_{j=k+1}^Ng_{jk}^\alpha, \qquad \sum_{i<j}g_{ij}^\alpha \delta^k_j = \sum_{j=1}^{k-1}g_{jk}^\alpha,
\end{equation}
and the derivative of the log likelihood with respect to $J$ is
\begin{equation}
\label{eq: mlp equation spin variables J}
\begin{split}
-\frac{\partial \mathcal{L}}{\partial J} = \sum_{i<j} \left( -\frac{4}{M} \sum_{\alpha < \beta} g_{ij}^\alpha g_{ij}^\beta + 2M\left(u_{ij}^*\right)^2\right)  = 0
\end{split}
\end{equation}
The maximum likelihood equations are therefore
\begin{equation}
\label{eq: MLP equation degrees}
\sum_{j\neq i}^N \sum_{\alpha=1}^M g_{ij}^\alpha = M \sum_{j\neq i}^N u_{ij}^*  \quad \forall i
\end{equation}
\begin{equation}
\label{eq: MLP equation overlap}
\frac{4}{M}\sum_{i<j}\sum_{\alpha<\beta}g_{ij}^\alpha g_{ij}^\beta = 2M \sum_{i<j}\left(u_{ij}^*\right)^2
\end{equation}
Note that the LHS of Equations \eqref{eq: MLP equation degrees} and \eqref{eq: MLP equation overlap} are precisely the quantities that we constrained from the start, namely $M\overline{k}_i$ and $4O/M$ respectively. According to the maximum likelihood principle, the empirical quantities $M\overline{k}_i$ and $4O/M$ equal their respective ensemble averages $M\langle \overline{k}_i \rangle _{\bm{\theta}}$ and $4\langle O\rangle_{\bm{\theta}}/M$. The quantity $u_{ij}$ can therefore be considered as an \textit{average} probability of a link occurring between the nodes $i$ and $j$ which is equal throughout the $M$ layers and is therefore a measure of the density of links in the multilink $m_{ij}$. This is similar to how we identified $p_{ij}$ to be the average link probability in the average layer configuration model, which was based solely on the constraints $\overline{k}_i$. In support of this idea, we see that in the case $J = 0$, 
\begin{equation}
\label{eq: u_ij zero J}
u_{ij}^*\Bigg|_{J = 0} = \frac{1}{2}\left(1 + \tanh{\left( -\frac{\theta_i + \theta_j}{2} \right)} \right) = \frac{e^{-(\theta_i + \theta_j)}}{1+e^{-(\theta_i + \theta_j)}}
\end{equation}
which is identical to the expression in Equation \eqref{eq: Example_3 Link Probability} which is the link probability $p_{ij}$ obtained in Section \ref{subsec: ACM} in the absence of the constraint for the overlap. The quantity $u_{ij}$ can therefore possibly be interpreted as a \textit{mean-field} quantity which \textit{globally} incorporates the layer interdependence that was introduced through the overlap $O$, but \textit{locally} treats the layers as if they are independent. A characteristic of mean field theories is that the effect of all elements of a system on a given element is approximated by a single averaged effect.

\section{Phase transitions in the AOCM}
\label{sec: Phase transitions}
\begin{figure*}[t]
\centering
\includegraphics[width = 1.0\textwidth]{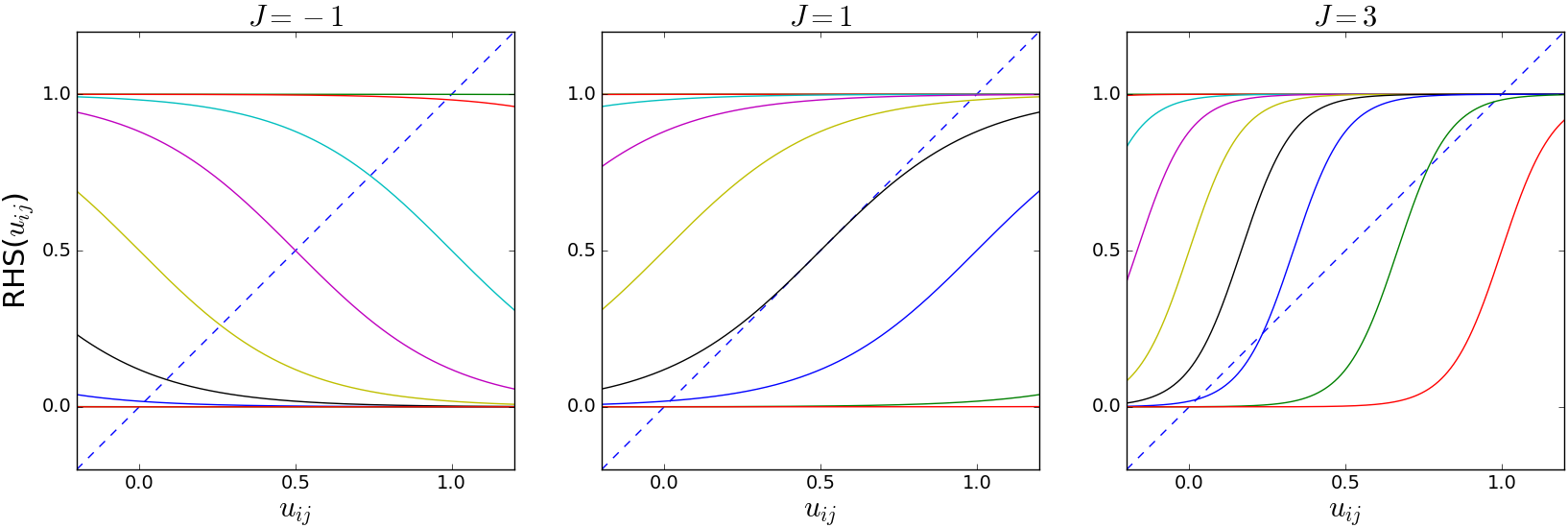} 
\caption[A graphical illustration of solution(s) to the self consistent equation]{A graphical illustration of the solution(s) of Equation \eqref{eq: u_ij tanh equation}. The $y$-axis shows the right hand side (RHS) of Equation \eqref{eq: u_ij tanh equation} while the $x$-axis shows the LHS. Each curve corresponds to a different choice of $\theta_{ij}$ where $\theta_{ij}\in \{-12,-8, -4, -2, 0, 2, 4, 8, 12 \}$ for the RHS while the dashed line corresponds to the LHS of Equation \eqref{eq: u_ij tanh equation}. The solutions of Equation \eqref{eq: u_ij tanh equation} are the intersections between the curves and the dashed line.}
\label{fig:tanh solutions} 
\end{figure*}

The amount of solutions $u_{ij}^* = u_{ij}(\theta_{ij},J)$ that Equation \eqref{eq: u_ij tanh equation} has depends on the values of the parameters $\theta_{ij}$ and $J$. We illustrate this fact in Figure \ref{fig:tanh solutions}, where both the LHS and the RHS of Equation \eqref{eq: u_ij tanh equation} are plotted for various values of $\theta_{ij}$ and $J$. 
The appearance of additional solutions introduce the possibility of \textit{phase transitions}, which are abrupt changes in the value of $u_{ij}^*$ and therefore also in the configuration and the properties of the multilink $m_{ij}$. These (vastly) different configurations of the multilink $m_{ij}$ that are separated by a phase transition are the \textit{phases} of the system $m_{ij}$. The point where additional solutions appear or vanish is called the \textit{bifurcation} point.

\begin{figure*}[t]
\centering
\includegraphics[scale=0.36]{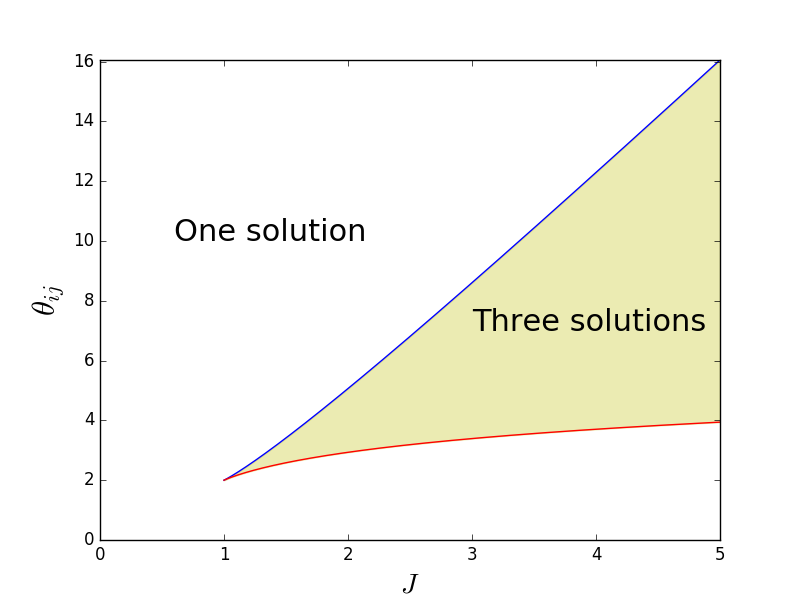} 
\caption[The solutions diagram for the AOCM]{The upper (blue) curve corresponds to the curve given in Equation \eqref{eq: theta_plus line}, which marks one of the branches of bifurcations of solutions $u_{ij}^*$ to Equation \eqref{eq: u_ij tanh equation}. The lower (red) curve corresponds to the bifurcation curve given in Equation \eqref{eq: theta_minus line}. In between the two curves (which is marked as the yellow area), there are three solutions $u_{ij}^*$ to Equation \eqref{eq: u_ij tanh equation}.}
\label{fig:solutions diagram} 
\end{figure*}

\begin{figure*}[t]
\centering
\includegraphics[width = 1.0\textwidth]{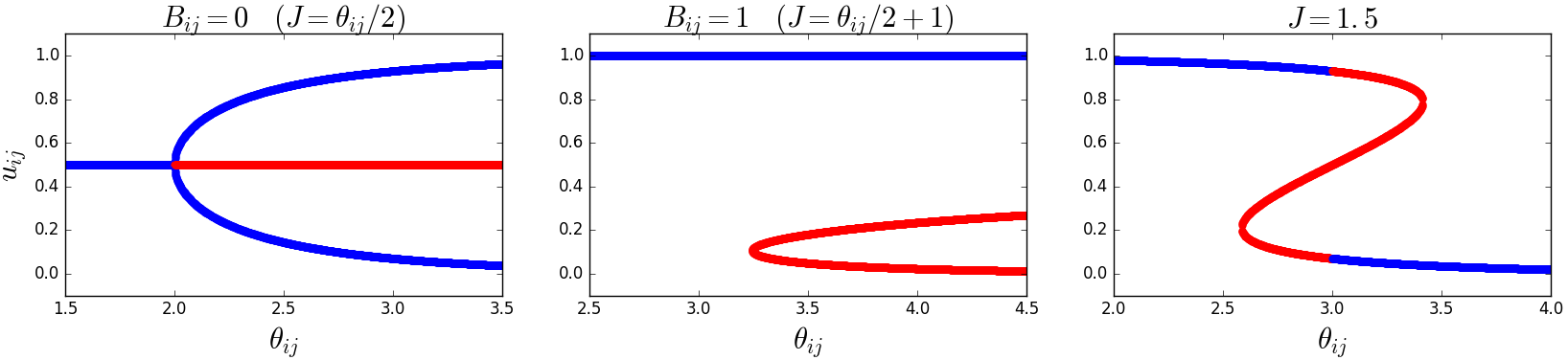} 
\caption[Stable and unstable solutions to the self consistent equation]{The blue segments of the curve(s) correspond to the stable solutions of Equation \eqref{eq: u_ij tanh equation} while the red segments of the curve(s) correspond to the unstable solutions. The left figure shows the solution $u_{ij}^*$ as a function of $\theta_{ij}$ while keeping $B_{ij}$ equal to zero. The middle figure shows the solution $u_{ij}^*$ as a function of $\theta_{ij}$ while keeping $B_{ij}$ equal to one. The right figure shows the solution $u_{ij}$ as a function of $\theta_{ij}$ for a constant value of $J = 1.5$, which translates to a non-constant $B_{ij}$.}
\label{fig:different field solutions} 
\end{figure*}

Figure \ref{fig:tanh solutions} shows that on the interval $0 \leq u_{ij} \leq 1$ there can be either 1, 2, or 3 solutions and that for $\theta_{ij} \rightarrow \infty$ or $\theta_{ij} \rightarrow -\infty$ there is always one solution $u_{ij}^* = 0$ and $u_{ij}^* = 1$ respectively. As can be seen in Figure \ref{fig:tanh solutions}, the number of solutions depends on whether the slope (derivative) of the RHS exceeds the slope of the LHS of Equation \eqref{eq: u_ij tanh equation} at their intersection. New solutions appear or vanish at the point where Equation \eqref{eq: u_ij tanh equation} is satisfied \textit{and} the derivative of the LHS and RHS of Equation \eqref{eq: u_ij tanh equation} are equal:
\begin{equation}
\label{eq: tanh derivative}
1 = J\left(1-\tanh^2{\left(2Ju_{ij} - \frac{\theta_{ij}}{2}\right)}\right)
\end{equation}
Equation \eqref{eq: tanh derivative} can not be satisfied if $J\leq 1$, since $0\leq \tanh^2(x) < 1$ for $x\in \mathbb{R}$, and therefore if $J\leq 1$ a phase transition is impossible and there is a unique solution for $u_{ij}$. When $J > 1$, Equation \eqref{eq: tanh derivative} gives us two potential solution branches $u_{ij,\pm}^* = \frac{1}{2} \pm \frac{1}{2} \sqrt{1-1/J}$ where we've used that $2u_{ij}^*-1 = \tanh{\left(2Ju_{ij}^* - \theta_{ij}/2 \right)}$. Equation \eqref{eq: u_ij tanh equation} can be written as $\theta_{ij} = 4Ju_{ij} - \log{\left(u_{ij}/(1-u_{ij}) \right)} $ by using the identity $\tanh^{-1}{x} = \frac{1}{2}\log{\left[\left( 1+ x \right)/ \left( 1 - x\right) \right]}$. By then substituting $u_{ij,\pm}^*$ into this expression for $\theta_{ij}$, we obtain the equations for the two curves in $(J, \theta_{ij})$ space that mark the points where additional solutions appear or vanish:
\begin{equation}
\label{eq: theta_plus line}
\theta_{ij}^{+}(J) = \frac{2\sqrt{J}}{\sqrt{J}- \sqrt{J-1}} - \log{\left(  \frac{\sqrt{J} + \sqrt{J-1}}{\sqrt{J} - \sqrt{J - 1}} \right)}
\end{equation}
\begin{equation}
\label{eq: theta_minus line}
\theta_{ij}^{-}(J) = \frac{2\sqrt{J}}{\sqrt{J} + \sqrt{J-1}} - \log{\left(  \frac{\sqrt{J} - \sqrt{J-1}}{\sqrt{J} + \sqrt{J - 1}} \right)}
\end{equation}
which are shown in Figure \ref{fig:solutions diagram}. In the region between the two curves, there are three solutions $u_{ij}^*$ to Equation \eqref{eq: u_ij tanh equation}. 
While certain values of $\theta_{ij},J$ may solve the maximum likelihood equations \eqref{eq: MLP equation degrees} and \eqref{eq: MLP equation overlap}, the corresponding solutions $u_{ij}^*$ to Equation \eqref{eq: u_ij tanh equation} may not necessarily maximize the likelihood and are therefore not 'valid' (or \textit{stable}). Once the values $\theta_{ij}$ and $J$ that solve the maximum likelihood equations are found, the likelihood corresponding to this set of values can be written as a function of the configuration of the graph (or the collection of configurations of the multilinks $m_{ij}$). The configuration that minimizes the (pair) Hamiltonian therefore maximizes the likelihood. As Figure \ref{fig:tanh solutions} suggests, in the regime where there are three solutions $u_{ij}$, one solution's value will be relatively high which corresponds to a relatively high density of links in $m_{ij}$, another solution's value will be relatively low which corresponds to a relatively low density of links in $m_{ij}$, and the last solution's value will be in between the other two and corresponds to a intermediate density of links in $m_{ij}$. By inspecting the (pair) Hamiltonian (Equation \eqref{eq: Curie Weiss Model}) in terms of the $\sigma_{ij}^\alpha = 2g_{ij}^\alpha-1$ variable, it becomes clear which of the three solutions $u_{ij}^*$ are viable (stable).  In the case where $B_{ij} = 0$, or equivalently when $\theta_{ij} = 2J$, the (pair) Hamiltonian is symmetric with respect to a change in sign: $\sigma_{ij}^\alpha \rightarrow -\sigma_{ij}^\alpha$, which means that the high and low density solutions are equal. The intermediate density solution however will result in a lower value for the Hamiltonian than the high and low density solutions. The viable (stable) solutions are in this case therefore the high and low density ones. In the case where $B_{ij} \neq 0$, it is clear that the high density solution minimizes the Hamiltonian when $B_{ij} > 0$ and maximizes it when $B_{ij}<0$. The low density solution minimizes the Hamiltonian when $B_{ij} < 0 $ and maximizes it when $B_{ij} > 0$. The intermediate solution will however never minimize the Hamiltonian when $B \neq 0$ and is therefore never viable (stable). From these considerations it becomes clear that a phase transition, which is a sudden change in $u_{ij}$ may only happen when we cross from a negative (positive) $B_{ij}$ to a positive (negative) $B_{ij}$ (when $J>1$). Figure \ref{fig:different field solutions} shows the symmetric stable solutions $u_{ij}$ in the case where $B_{ij} = 0$, with the bifurcation occurring at $J = 1$. In case of the positive field $B_{ij} = 1$ it shows a single stable solution curve which is the high density solution (in the case where $B_{ij} = -1$ this image would be flipped with respect to the $u_{ij}^* = 1/2$ axis). The right figure in Figure \ref{fig:different field solutions} shows that the value of the stable solution $u_{ij}$ jumps when $B_{ij}$ crosses from positive to negative, as expected. 

\section{Numerical Analysis}
\label{sec: Numerical Analysis}
In the previous chapter we defined the AOCM and derived Equations \eqref{eq: u_ij tanh equation}, \eqref{eq: MLP equation degrees} and \eqref{eq: MLP equation overlap}, which are the principal equations of the model. This system of equations is generally however very difficult to solve, both analytically and numerically. Instead of creating a null model by solving the maximum likelihood equations to obtain correct values for the Lagrange multipliers, we shall treat the Lagrange multipliers as free parameters in this chapter in order to explore and analyze the properties of the system as a function of these parameters. This analysis shall be done by utilizing various numerical methods, such as the Metropolis-Hastings algorithm \cite{hastings1970monte}. This algorithm can be used to sample the exponential probability distribution which is defined by the Hamiltonian of the model. By sampling the distribution we may obtain various properties of the graph ensemble in a numerical manner which is independent of our analytical results. These numerical results may then be compared to our analytical results in order to test the validity of the analytical results. Note that the sampling of the exponential distribution defined by a specific Hamiltonian may also be referred to as the simulation of a multiplex that corresponds to that Hamiltonian.

\subsection{Exploring the parameter space}
The functional form of the statistical distributions characterizing large networks generally defines two broad network classes. The first refers to the so-called statistically \textit{homogeneous} networks where the distribution that characterizes (for example) the degree has functional forms with fast decaying tails such as Gaussian or Poisson distributions. The second class refers to statistically \textit{heterogeneous} networks where the distribution that characterizes various measures such as the degree corresponds to heavy (or "fat") tailed distributions. We will explore the parameter space $(\theta_1,\ldots,\theta_N,J)$ of the model by specifying a value for $J$ and sampling $x_1,\ldots,x_N$ from a distribution for each class where
\begin{equation}
x_i \equiv e^{-\theta_i}.
\end{equation}
The reason for sampling $x_i$ instead of $\theta_i$ from a chosen distribution is that the chosen parameters are then easily relatable to the parameters in the Configuration Model which is often studied by specifying distributions from which $x_i$ is sampled as well.

\subsubsection{Constant $x_i$: Erd\H{o}s–-R\'enyi graphs with overlap}
\label{subsec: delta distribution}
One possible distribution that corresponds to the class of statistically homogeneous networks from which we can sample $x_1,\ldots, x_N$ is the constant case where $x_1 = x_2 = \ldots = x_N = x$ and therefore $\theta_1 = \theta_2 = \ldots = \theta_N = \theta$ all have the same value $x$ and $\theta$ respectively. In this case the chosen distribution is essentially a Delta distribution, which is sharply peaked. With this choice of the parameters, our model is an extension of the Erd\H{o}s–-R\'enyi model, which is a random graph model where all of the links in the entire graph occur with the same probability and this model can be derived by solely constraining the total number of links in the network within the ERGM.

By looking at Equation \eqref{eq: Hamiltonian}, we can see that a uniform $\theta$ essentially means that instead of constraining the average layer degrees $\overline{k}_i$, we constrain the total number of links $L$ in the multiplex network. In this case, the equations become

\begin{equation}
\label{eq: u_ij tanh equation uniform theta}
u = \frac{1}{2} + \frac{1}{2}\tanh{\left( 2Ju - \theta\right)}
\end{equation}
\begin{equation}
\label{eq: MLP equation links uniform theta}
\sum_{i<j}^N \sum_{\alpha=1}^M g_{ij}^\alpha = \frac{MN(N-1)}{2} u^* = \langle L \rangle 
\end{equation}
\begin{equation}
\label{eq: MLP equation overlap uniform theta}
\frac{4}{M}\sum_{i<j}\sum_{\alpha<\beta}g_{ij}^\alpha g_{ij}^\beta = MN(N-1)\left(u^*\right)^2 = \frac{4}{M}\langle O \rangle 
\end{equation}
where $u^* = u(\theta, J)$ is the solution to Equation \eqref{eq: u_ij tanh equation uniform theta}. Note that we now have a single equation for $u$, which means that there is a possibility of a single \textit{global} phase transition across the multiplex network instead of the possibility of independent phase transitions for every multilink $m_{ij}$. Additionally, we note that if $u^*$ can be considered as the density (and the link probability) of the network, the value of $u^*$ is exactly the same as the value of the density $p$ in the Erd\H{o}s–-R\'enyi model  \cite{erdos1960evolution}, \cite{park2004statistical} which solely constrains the number of links in the network. The difference between our model and the Erd\H{o}s–-R\'enyi model is that our model contains the possibility of a phase transition. An unfortunate similarity between these two models is the fact that the number of links $\langle L \rangle$ determines the overlap $\langle O \rangle$, which means that according to the analytical results they can not be tuned independently from each other. 

\begin{figure*}[t]
\centering
\includegraphics[width = 1.0\textwidth]{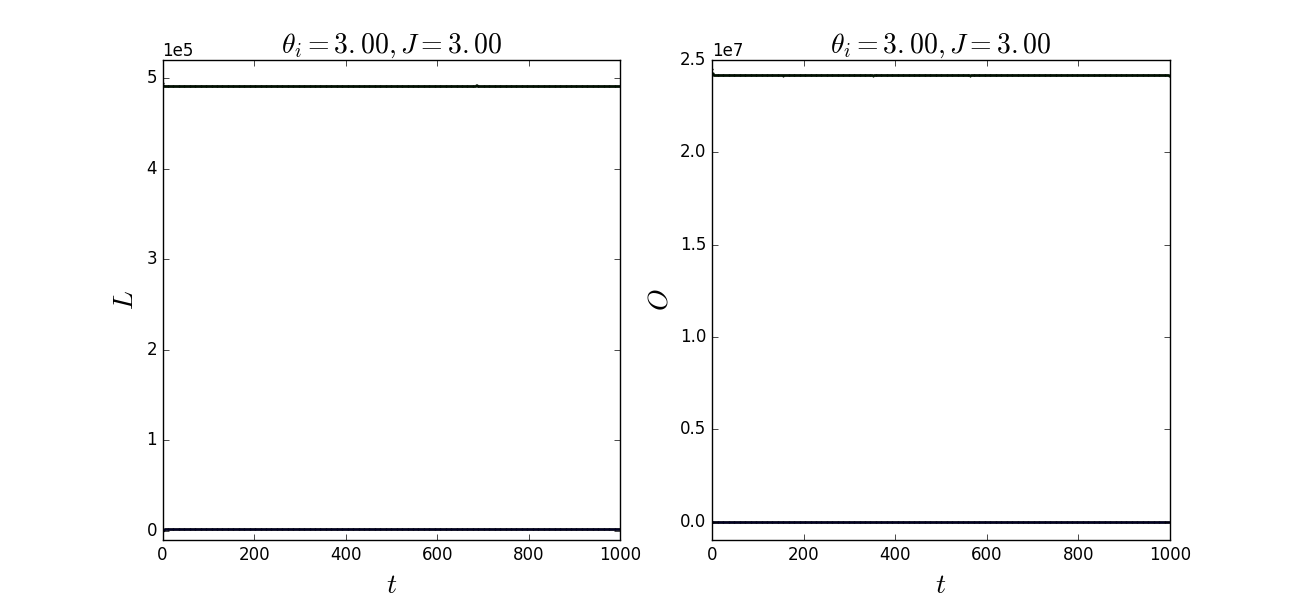} 
\caption[Symmetric convergence of the Metropolis-Hastings algorithm]{The left figure illustrates the total number of links $L$ in the multiplex network as a function of the time during the sampling of the exponential distribution using two different initial configurations. The right figure illustrates the overlap $O$ in the multiplex network as a function of the time using two different initial configurations. The initial configurations (at $t=0$) are fully connected multiplexes and completely empty multiplexes which finally converge to multiplexes with vastly different densities. Every time iteration $\Delta t$ actually represents $\Delta t = MN(N-1)/2$ metropolis iterations, which is equal to the total number of possible links in the multiplex network. In this particular simulation we have used the values $M = 100$, $N = 100$, $\theta = 3.0$,  and $J = 3.0$.}
\label{fig:symmetry} 
\end{figure*}

By using the Metropolis-Hastings algorithm, we have sampled the exponential random graph ensemble for multiplexes with $M = 100$ layers and $N = 100$ nodes for various values of $\theta$ and/or $J$. 
\begin{figure*}[t]
\centering
\includegraphics[width = 1.0\textwidth]{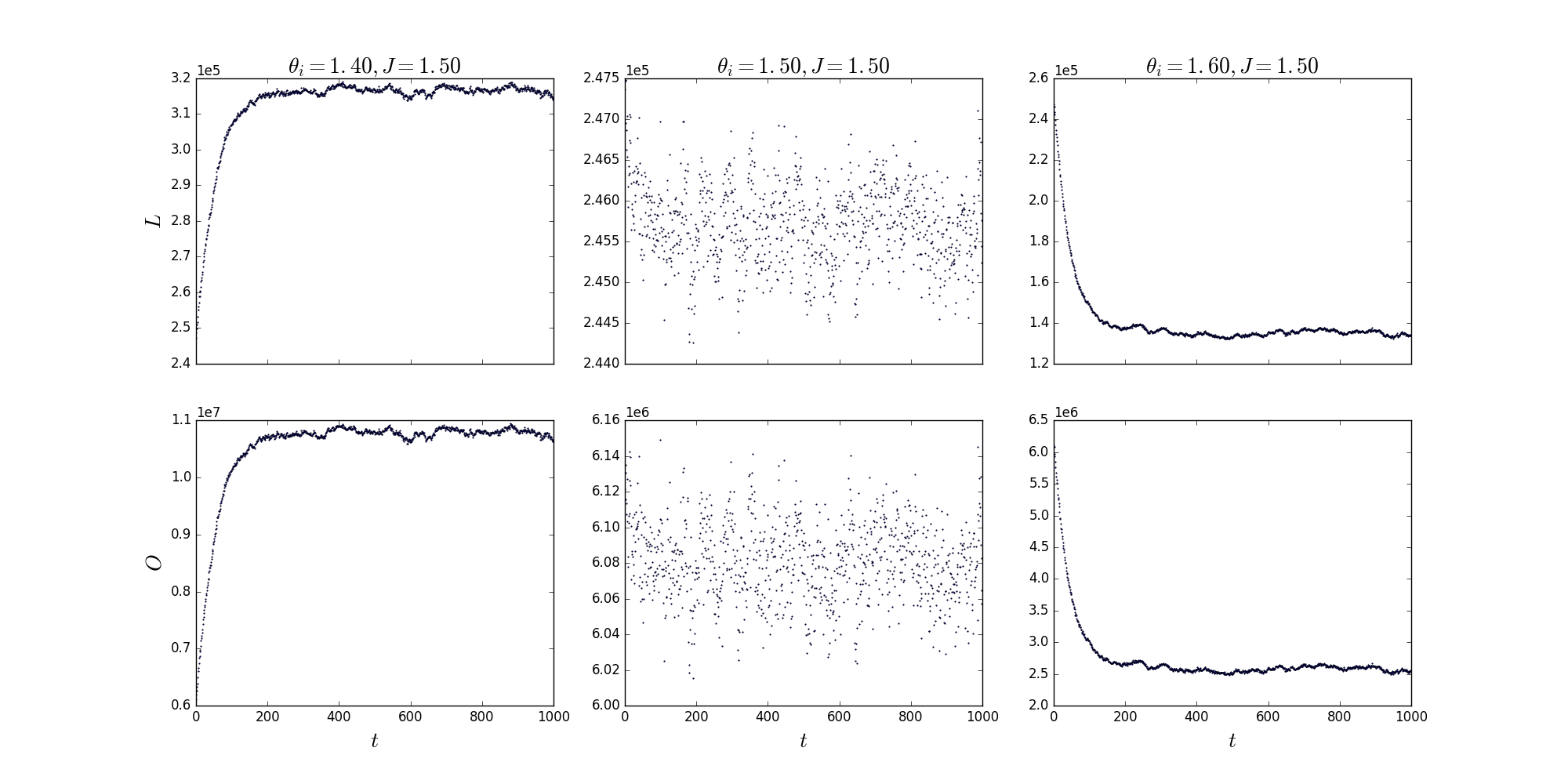}
\caption[Numerical simulation of the phase transition]{The upper three plots illustrate the total number of links $L$ in the multiplex network as a function of the simulation time. The lower three plots illustrate the overlap $O$ in the multiplex network as a function of the simulation time. The left plots correspond to a simulation where $\theta = 1.4$, the middle plots correspond to a simulation where $\theta = 1.5$, and the right plots correspond to a simulation where $\theta = 1.6$. Every simulation is done with the values $J = 1.5$, $N = 100$, and $M = 100$. This collection of figures shows that there is a phase transition from a high density phase to a low density phase.}
\label{fig:phase transition simulation} 
\end{figure*}
If we repeat the simulations for $J = 1.5$ and $\theta = 1.4$, $\theta = 1.5$, and $\theta = 1.6$, the system must undergo a phase transition according to the right figure in Figure \ref{fig:different field solutions} in Section \ref{sec: Phase transitions} at $ J = \theta = 1.5$. We expect an abrupt change in the value of $u^*$ and according to Equations \eqref{eq: MLP equation links uniform theta} and \eqref{eq: MLP equation overlap uniform theta} we therefore expect an abrupt change in the equilibrium value of both $L$ and $O$. Figure \ref{fig:phase transition simulation} shows these simulations for $\theta \in \{1.4, 1.5, 1.6\}$. This figure clearly shows the transition from a relatively high density multiplex to a low density multiplex once the value of the field $B = J - \theta$ changes sign. These simulations have been repeated for different combinations of values for $J$ and $\theta$ around the point where $B$ changes sign. These results have been qualitatively similar and will therefore not be shown here. Note that the middle plot in Figure \ref{fig:phase transition simulation} shows that the algorithm converges to multiplexes with a density of $1/2$, which means that $L$ is approximately half of the total amount of possible links in the multiplex. However, we expect the algorithm to converge to a low density or a high density multiplex configuration with equal probability with the mean value of the density being $1/2$, as can be seen in the left plot of Figure \ref{fig:different field solutions}. This is likely the result of the value of $J$ being too "small", which means that the interactions between the layers of the multiplex are not strong enough. The relatively weak interaction likely results in the system acting as if there is no interaction at all and the stable solution is in that case the intermediate valued solution shown in the left plot of Figure \ref{fig:different field solutions}. To support this claim, we repeat the simulation for a higher value of $J$, namely $J = 3$, and $\theta = J = 3$ while omitting the simulations with a value of $\theta$ in the vicinity of $\theta = J$, since qualitatively the plots of these simulations remain identical to the ones shown on the left and on the right in Figure \ref{fig:phase transition simulation}. The resulting plots are given in Figure \ref{fig:symmetry}, which clearly show the convergence of the simulation to multiplexes of either extremely high or extremely low densities. Note that in these plots we have used an initially complete and empty multiplex configuration. This was also done in the case of $J=1.5$, however the simulation then still converged to a multiplex configuration with a density of $1/2$ instead of a configuration with an extremely high or low density.

\begin{figure*}
\begin{center}
\includegraphics[width=.49\textwidth]{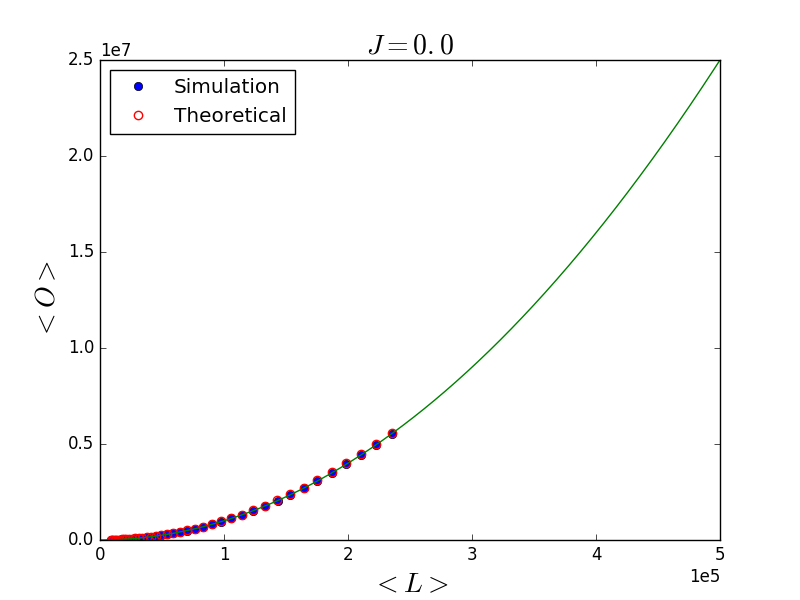}
\includegraphics[width=.49\textwidth]{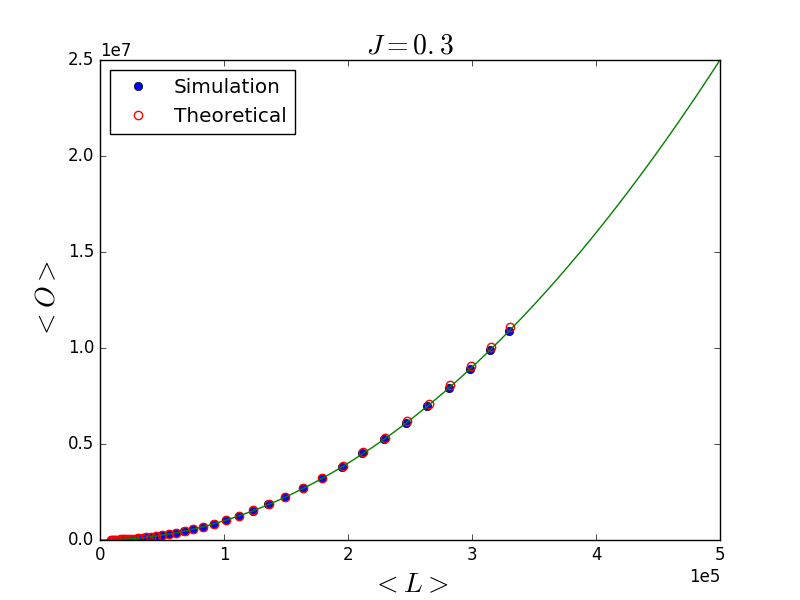}
\includegraphics[width=.49\textwidth]{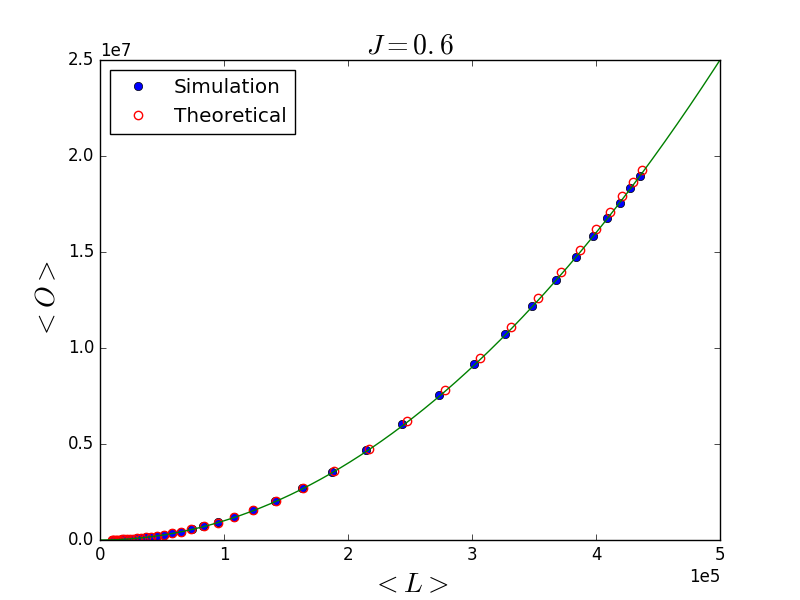}
\includegraphics[width=.49\textwidth]{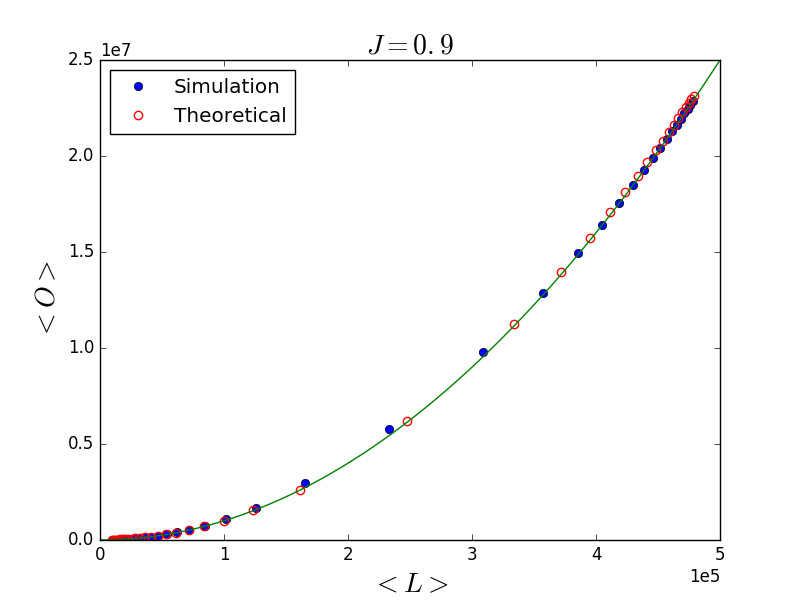}
\includegraphics[width=.49\textwidth]{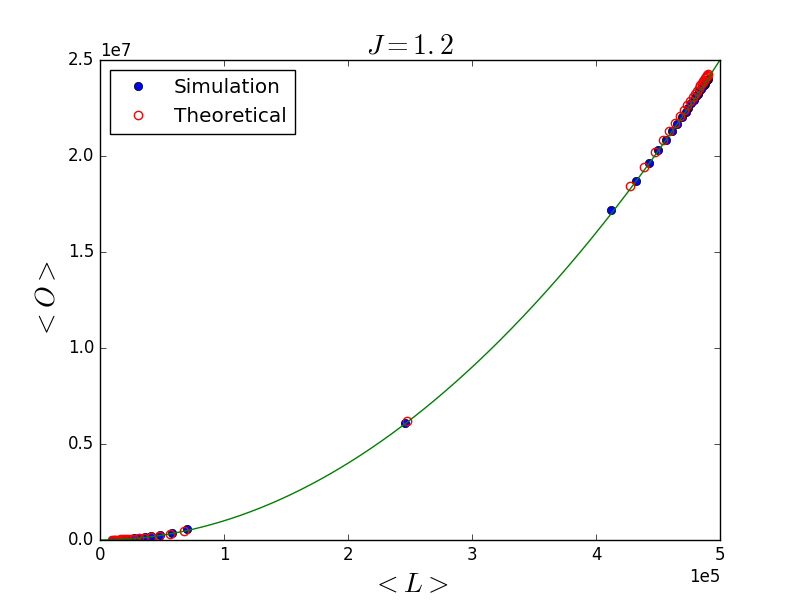}
\includegraphics[width=.49\textwidth]{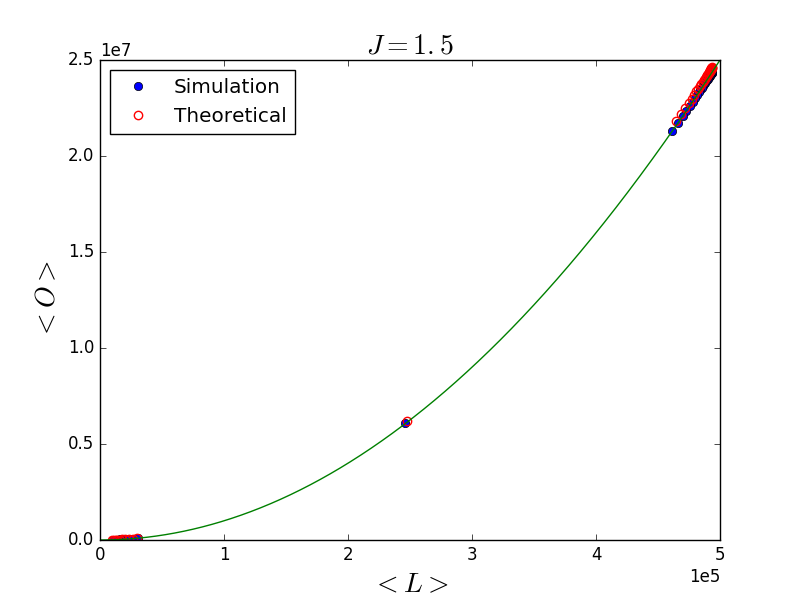}
\end{center}
\caption{The blue points correspond to the mean values of $O$ and $L$ which are obtained by using the Metropolis-Hastings algorithm for $J\in \{0.0, 0.3, 0.6, 0.9, 1.2, 1.5\}$ and a $\theta \in [0.05, 2.00]$ in steps of $\Delta \theta = 0.05$ in the case where $x_i$ is constant. The red open circles are the theoretically predicted points corresponding to the same $\theta$ and $J$ that are used in the simulations. The green curve corresponds to the curve $\langle O \rangle = \langle L \rangle ^2 / N^2$. Multiple solutions for $u_{ij}^*$ first appear when $J > 1$.}
\label{fig:O L relationship various J}
\end{figure*}

Equations \eqref{eq: MLP equation links uniform theta} and \eqref{eq: MLP equation overlap uniform theta} predict a quadratic relationship between $O$ and $L$: $\langle O \rangle = \langle L \rangle^2 /N^2$. In order to verify this relationship numerically, we again simulate multiplexes with $M = 100$ layers, $N = 100$ nodes, and a variety of values for $\theta$ and $J$. Each simulation results in a value for $\langle L \rangle$ and a value for $\langle O \rangle$ which we plot against each other. These points are compared to the theoretical points that are predicted by equations \eqref{eq: u_ij tanh equation uniform theta}, \eqref{eq: MLP equation links uniform theta} and \eqref{eq: MLP equation overlap uniform theta} for the chosen parameter values which lie on the line $\langle O \rangle = \langle L \rangle ^2 / N^2$. The result are shown in Figure \ref{fig:O L relationship various J}. The figure shows that the relationship between simulated quantities is in agreement with the relationship between the quantities that is predicted by the model and that the actual theoretical predictions are quite accurate for this particular choice of the parameter values.

Figure \ref{fig:solutions diagram} indicates that phase transitions first become possible when $J > 1$. In order to demonstrate the existence of a phase transition the simulations are done for different values of $J$ which results in Figure \ref{fig:O L relationship various J}. From this figure it is apparent that when $J$ approaches $1$ from below, both the simulation data and the corresponding theoretically predicted data diverge from intermediate values of $\langle L \rangle$ towards low and high values of $\langle L \rangle$. When $J>1$ the multiplex networks corresponding to the data points are either in the very low density or the very high density case, which is an indication of a phase transition occurring when increasing the value of $J$ in this case. Another notable observation that can be made is the increased inaccuracy of the theoretical predictions for data points that correspond to intermediate values of $\langle L \rangle$. These intermediate points correspond to parameter values where $\theta \approx J$ which means that $B_{ij} = J - \theta  \approx 0$. A possible explanation for the discrepancy between the theoretical results and the simulation data is that the configuration that the Metropolis-Hastings algorithm converges to is the most typical configuration for the given Hamiltonian but is not necessarily representative of the configuration that corresponds to the ensemble average. When $B_{ij} \approx 0$ the Hamiltonian has two local minima of comparable values, one of which is the global minimum. The Metropolis-Hastings algorithm may then converge to either local minimum with comparable probabilities. The value of the theoretical prediction corresponds to the \textit{ensemble average}, which actually lies between the values corresponding to the two local minima. When $B_{ij}$ is not approximately $0$ there are still two local minima, however the global minimum is much more likely to occur than the local mimimum which is not the global minimum. This means that in this case, the ensemble average of a quantity is approximately the same as the value of the quantity for the most typical/likely multiplex configuration to occur (which results in accurate predictions) while in the case where $B_{ij}\approx 0$ the most typical configuration does not accurately represent the configuration that corresponds to the ensemble average.

\subsubsection{Power law distribution of $x_i$: scale-free networks with overlap}
\label{subsec: power law distribution}
Many real-world networks contain a high level of statistical heterogeneity. The majority of the vertices of these real-world networks have a small number of links to other vertices while a few vertices have a relatively high number of links to other vertices, which are also referred to as "hubs". An example is the World Wide Web where some pages are incredibly popular and are pointed to by thousands of other pages, while generally most pages are almost unknown. The presence of these hubs in a network often results in a degree distribution $P(k)$ with heavy tails \cite{barabasi1999emergence}~, where the degrees vary over a broad range, often spanning several orders of magnitude. This heavy tail can be approximated by a power-law distribution $P(k) \sim k^{-\gamma}$. In heavy tail degree distributions, vertices with a degree much larger than the average degree $\langle k \rangle $ occur with a non-negligible probability. 

In the Configuration Model (see subsection \ref{subsec: ACM}) the expected degree distribution is determined by the hidden variables $\theta_i$ or equivalently the transformed hidden variables $x_i = e^{-\theta_i}$. If $x$ is distributed according to a power-law, the expected degree distribution shall be distributed according to a power-law as well. Networks with a power-law degree distribution are also referred to as scale-free networks. Since our model is an extension of the Configuration Model, we find it a suitable first choice to sample  $x_ i = e^{-\theta_i}$ from a power-law distribution $P(x) \sim x^{-\gamma}$ for various values of $\gamma$ even though the expected degree distribution is not solely determined by the hidden variables $x_i$ (or $\theta_i$) but likely depends on $J$ as well. However, a higher level of heterogeneity in the hidden variables $x_i$ will lead to a higher level of heterogeneity in the degrees. Since the parameter space is quite large ($N+1$ dimensional) we define 
\begin{equation}
x_i = z x_{0,i}
\end{equation}
where $z$ is a scaling factor. We sample $x_{0,i}$ only \textit{once} from every chosen distribution. The value of $x_i$ is varied by varying the scaling factor $z$. The parameter space to be explored will then be $(z,J)$ which is $2$ dimensional. We have that 
\begin{equation}
\theta_i = -\log{\left(zx_{0,i}\right)}
\end{equation}
which shows that an increasing $z$ leads to a decreasing $\theta_i$. In the Configuration Model the link probability is equal to $p_{ij} = x_ix_j/(1+x_ix_j)$ which means that generally a decreasing $\theta$ leads to an increasing number of links (or the density) in the network. This relationship between $\theta$ and the number of links holds in our model as well. 

\begin{figure*}
\centering
\includegraphics[width = 0.49\textwidth]{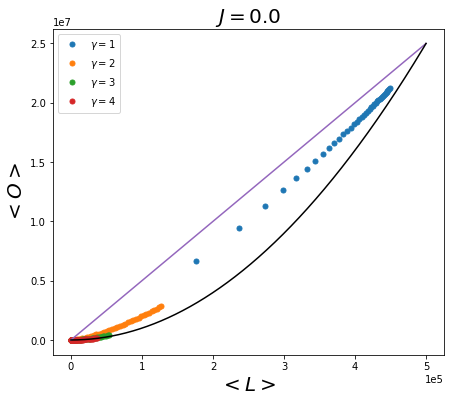}
\includegraphics[width = 0.49\textwidth]{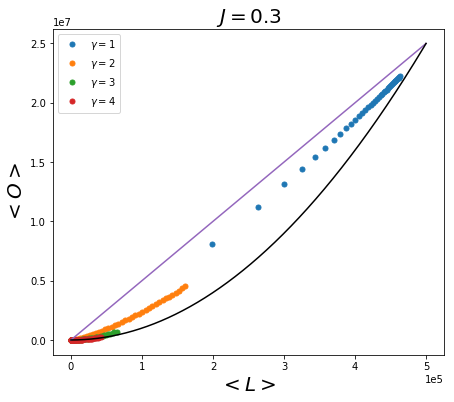}
\includegraphics[width = 0.49\textwidth]{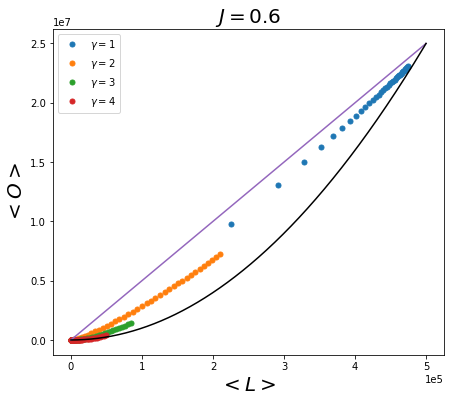}
\includegraphics[width = 0.49\textwidth]{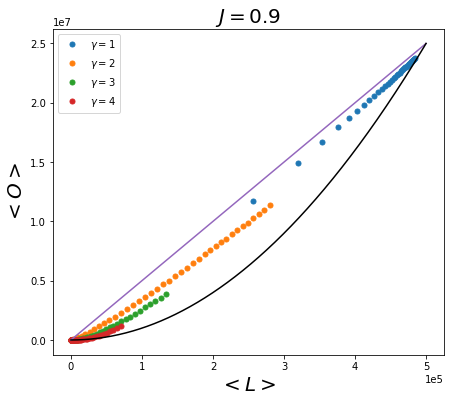}
\includegraphics[width = 0.49\textwidth]{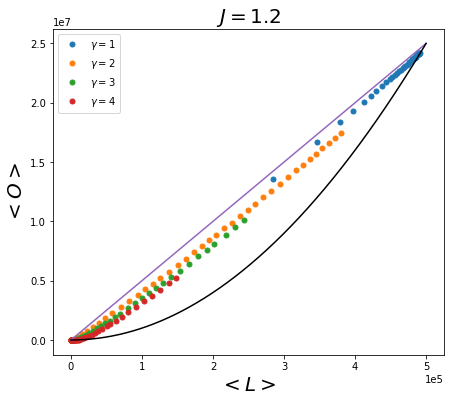}
\includegraphics[width = 0.49\textwidth]{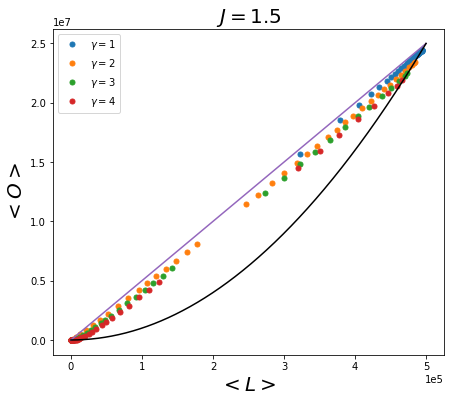}
\vspace{-10pt}
\caption[Short]{The blue, green, red and the yellow colored points correspond to the mean values of $O$ and $L$ which are obtained by using the Metropolis-Hastings algorithm for $J \in \{0.0, 0.3, 0.6, 0.9, 1.2, 1.5 \}$ and a $z \in [0.05, 2.00]$ in steps of $\Delta z = 0.05$ where $x_{0,i}$ is sampled from a power law distribution with different values for $\gamma$ as indicated in the legend. The purple solid line corresponds to the line $\langle O \rangle = \frac{M}{2}\langle L \rangle $. The black solid line corresponds to the curve $\langle O \rangle = \langle L \rangle ^2 / N^2 $. For this plot we've chosen that $M = 100$ and $N = 100$.}
\label{fig:CM powers comparison} 
\end{figure*}

The complexity of Equations \eqref{eq: MLP equation degrees}, \eqref{eq: MLP equation overlap} and \eqref{eq: u_ij tanh equation} does not allow us to easily derive the expected relationship between the overlap and the number of links in the network, as was the case when $\theta_i$ was constant. It is however possible to visualize the relationship between the overlap and the number of links by using the Metropolis-Hastings algorithm. Figure \ref{fig:CM powers comparison} shows this relationship where $x_i$ is sampled from power law distributions with various values of $\gamma$ in addition to the data points corresponding to the case where $x_i$ was sampled from a delta distribution (see the previous subsection) for comparison. This figure shows that the overlap for a given number of links is higher in the cases where $x$ is drawn from a power law distribution than when $x$ is drawn from a Delta distribution even though the coupling parameter $J$ is kept constant. The black curve in Figure \ref{fig:CM powers comparison} is the quadratic curve along which the data points sampled from the Delta distribution lie. The cause of this difference lies in the level of heterogeneity of the chosen distributions. A more heterogeneous distribution, such as the power law distribution, leads to a small number of nodes $i$ having a large corresponding value for $x_i$ since large values of $x_i$ occur with a non-negligible probability. As we had established earlier, our model is an extension of the average layer configuration model where the layers of the multiplex are treated equally and as if they are independent. This means that a high value of $x_i$ results in node $i$ being a hub in every layer of the multiplex, which means that the degree $k_i^\alpha$ of node $i$  in layer $\alpha$ is generally high for every value of $\alpha$. The presence of these large hubs therefore lead to an increased overlap compared to the case where $x_i$ is drawn from a Delta distribution due to the increased heterogeneity of the network, and not an increased coupling between layers. This effect can also be seen in Figure \ref{fig:CM powers comparison} when comparing the results of the different power law distributions with each other. The figure shows that a smaller value of $\gamma$ leads to a higher overlap for a given number of links. By increasing the value of $\gamma$ the power law distribution shall become more sharply peaked and will therefore be more similar to the Delta distribution. Note however that increasing the value of the coupling parameter $J$ itself also leads to an increase in the overlap for a given number of links for the same distribution.

\begin{figure*}[!htb]
\begin{center}
\includegraphics[width=.49\textwidth]{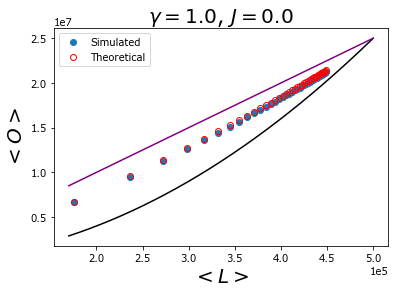}
\includegraphics[width=.49\textwidth]{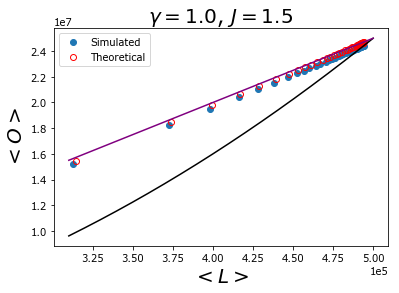}
\end{center}
\caption{The blue points correspond to the mean values of $O$ and $L$ which are obtained by using the Metropolis-Hastings algorithm for $J = 0$ in the left plot and $J = 1.5$ in the right plot and every data point corresponds to a $z \in [0.05, 2.00]$ in steps of $\Delta z = 0.05$ in the case where $x_{i,0}$ is drawn from a power law distribution with $\gamma = 1$. The red open circles are the theoretically predicted points corresponding to the same $x_i$ and $J$ that are used in the simulations. The purple solid line corresponds to the line $\langle O \rangle = \frac{M}{2}\langle L \rangle $. The black solid line corresponds to the curve $\langle O \rangle = \langle L \rangle ^2 / N^2 $. For these plot's we've chosen that $M = 100$ and $N = 100$.}
\label{fig:O L relationship CM}
\end{figure*}

By using equations \eqref{eq: u_ij tanh equation}, \eqref{eq: MLP equation degrees} and \eqref{eq: MLP equation overlap} we calculate the theoretically predicted values of $\langle O \rangle$ and $\langle L \rangle $ and compare them to the simulation data where $x_{0,i}$ is sampled from a power law distribution with $\gamma = 1$. The result is shown in \ref{fig:O L relationship CM}. The results for $\gamma \in \{2,3,4\}$ are qualitatively similar and are therefore not shown here. Figure \ref{fig:O L relationship CM} shows that the theoretical predictions are in good agreement with the simulation data. A notable observation is that when we compare Figure \ref{fig:O L relationship CM} to Figure \ref{fig:O L relationship various J} the relationship between $\langle O \rangle $ and $\langle L \rangle$ now seems to be linear instead of quadratic. This observation suggests that the general equations \eqref{eq: u_ij tanh equation uniform theta}, \eqref{eq: MLP equation degrees} and \eqref{eq: MLP equation overlap} can not be solved analytically in a way that provides an explicit relationship between $\langle O \rangle$ and $\langle L \rangle $ without specifying the values of $\theta_i$ first. The approach of the $(\langle L \rangle, \langle O \rangle)$ curve towards linear behavior is likely due to the multiplex configurations approaching maximum overlap for a given number of links. In case of maximum overlap for a given number of links, we have that $g_{ij}^\alpha = g_{ij}^\beta = g_{ij}$ for every $(\alpha,\beta) \in \{1,\ldots,M\}$. This leads to 
\begin{equation}
\begin{split}
O = \sum_{\alpha < \beta } \sum_{i<j} g_{ij}^\alpha g_{ij}^\beta = \sum_{\alpha<\beta }\sum_{i<j}\left(g_{ij}^\alpha\right)^2 = \sum_{\alpha<\beta}\sum_{i<j} g_{ij} = \frac{M}{2} L.
\end{split}
\end{equation}
This curve (or relationship), which signifies the theoretical maximum overlap for a given number of links, may also be referred to as the \textit{linear upper limit} and it is plotted in Figure \ref{fig:CM powers comparison}. Figure \ref{fig:CM powers comparison} shows that a combination of the heterogeneity due to the power law distribution and the coupling due to $J$ leads to an overlap that is almost maximal for a given number of links in the network. Figures \ref{fig:O L relationship CM} and \ref{fig:CM powers comparison} show that for a lower values of $J$ the $(\langle L \rangle , \langle O \rangle )$ line is slightly steeper than the line $\langle O \rangle = M/2 \langle L \rangle $, which means that while the overlap in this case is very high the heterogeneity due to the power law distribution in absence of coupling due to $J$ is not sufficient to create a maximally overlapping network.

In the case of a constant $x_{i}$, which we obtain by sampling $x_{i}$ (or $x_{0,i}$) from a delta distribution, we had shown that the relationship between the overlap and the number of links is quadratic. We may refer to this specific quadratic relationship between the overlap and the number of links as the \textit{quadratic lower limit} which is plotted in Figures \ref{fig:CM powers comparison} and \ref{fig:O L relationship CM}. For a given number of links, the overlap was generally much smaller than in the case where $x_{0,i}$ was sampled from a power law distribution.

\subsubsection{Log-normal distribution of $x_i$}
A log-normal distribution is a continuous probability distribution of a random variable whose logarithm is normally distributed. Thus, if the random variable $x$ is log-normally distributed, then $y = \ln{x}$ has a normal distribution. The probability density for a log-normal distribution is 
\begin{equation}
P(x) = \frac{1}{x} \frac{1}{\sigma \sqrt{2\pi}}e^{-\left(\ln{x}-\mu\right)^2/\left( 2\sigma \right)}
\end{equation}
where $\mu$ and $\sigma$ correspond to the mean and the standard deviation of the normal distribution of $\ln{x}$. Analogous to the method used in Subsection \ref{subsec: power law distribution} to explore the parameter space, the value of $x_{i}$ is again varied by introducing a scaling factor that can be varied such that $x_i = z x_{0,i}$ and $\theta_i = -\log{\left( z x_{0,i} \right)}$ where we sample $x_{0,i}$ \textit{once} from the log-normal distribution for a variety of values for $\mu$ and $\sigma$.

The log-normal distribution has an interesting property that allows us to see the transition in the relationship between the overlap and the number of links from the quadratic lower limit to the linear upper limit by varying the value of $\sigma$. When $0 < \sigma \ll 1$ the normal distribution of $\ln{x_{0,i}}$ becomes sharply peaked. By decreasing the value of $\sigma$ towards $0$,  $\ln{x_{0,i}}$ (and therefore $x_{0,i}$ as well) shall be distributed according to a delta distribution. This was the distribution that was used in Subsection \ref{subsec: delta distribution} which led to the quadratic lower limit relationship between the overlap and the number of links in the network. When $\sigma \geqq 1$, the log normal distribution converges to a power law distribution with $\gamma = 1$. This was (one of) the distribution(s) that was used in Subsection \ref{subsec: power law distribution} which led to the linear upper limit relationship between the overlap and the number of links in the network (when $J$ was sufficiently large). By increasing the value of $\sigma$ from $0$ to a sufficiently large value (e.g. $\sigma = 10$) we therefore increase the heterogeneity of the network starting from a completely homogeneous network ($\sigma \approx 0$) and observe the transition from the quadratic lower limit relationship between the overlap and the number of links to the linear upper limit relationship in the simulation data. 

\begin{figure*}
\begin{center}
\includegraphics[width=.49\textwidth]{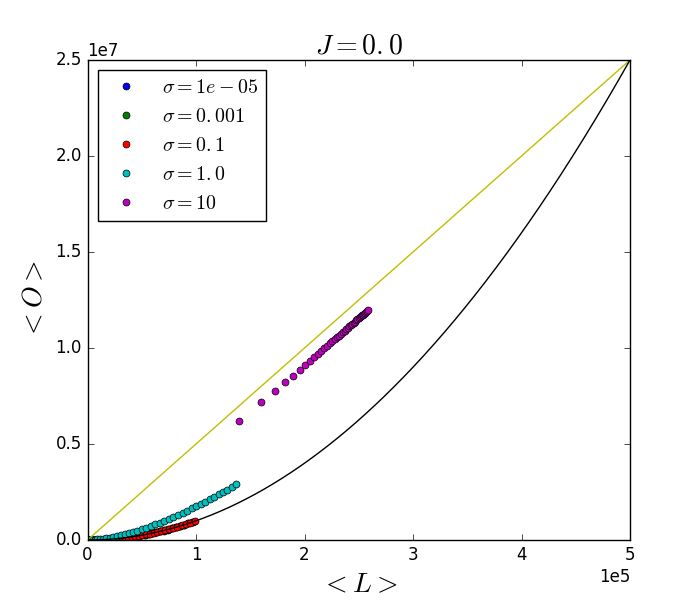}
\includegraphics[width=.49\textwidth]{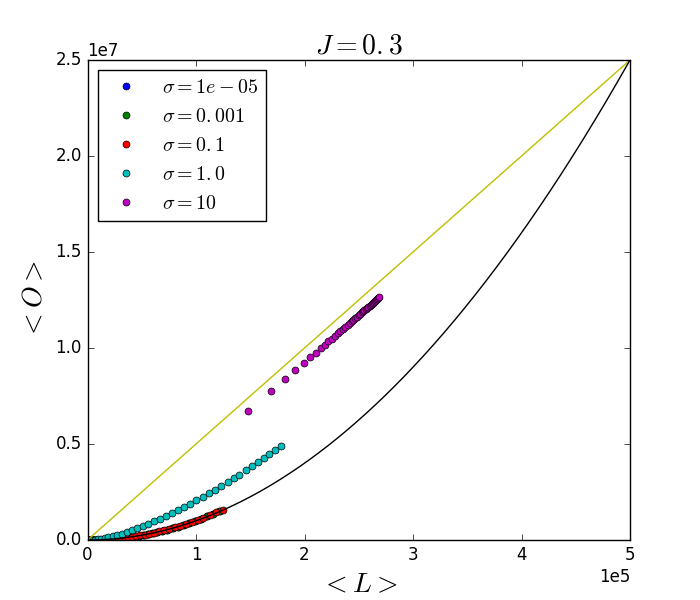}
\includegraphics[width=.49\textwidth]{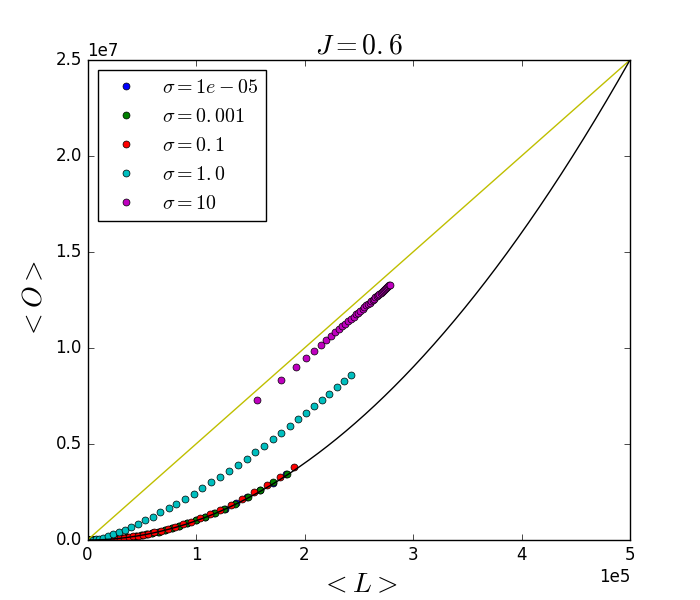}
\includegraphics[width=.49\textwidth]{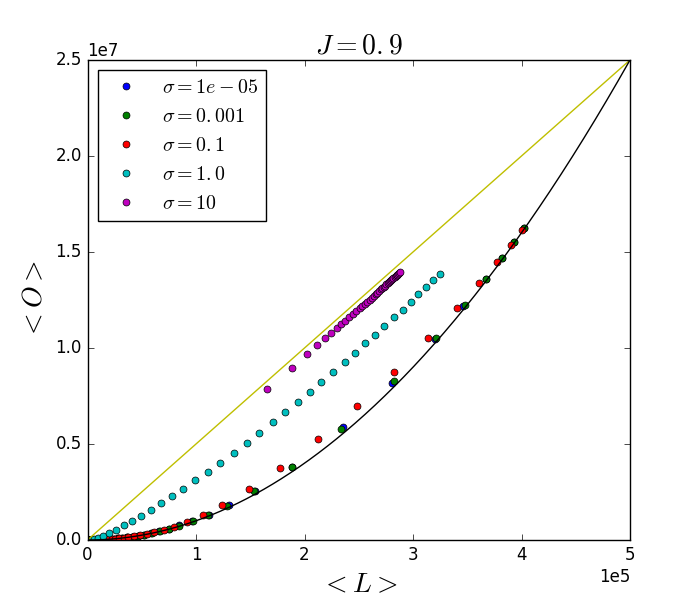}
\includegraphics[width=.49\textwidth]{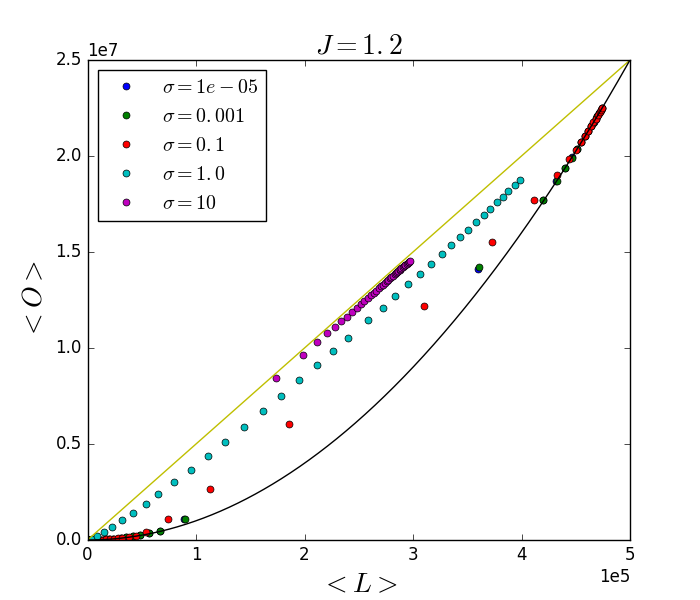}
\includegraphics[width=.49\textwidth]{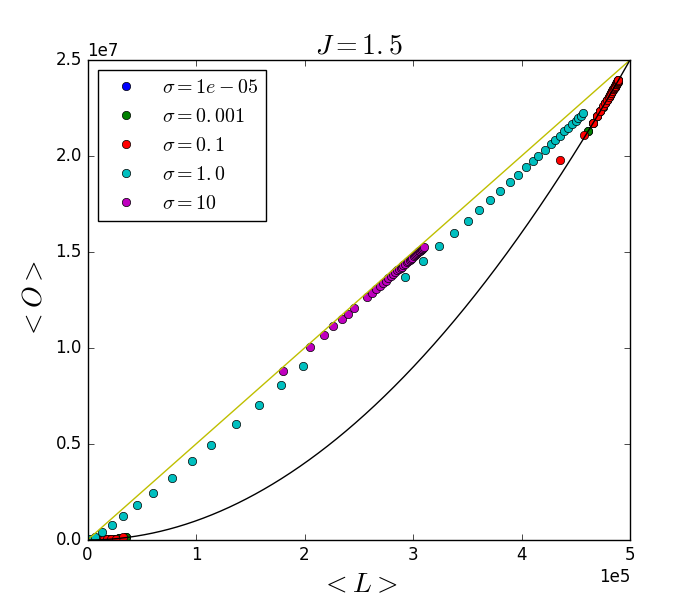}
\end{center}
\vspace{-20pt}
\caption{The colored points correspond to the mean values of $O$ and $L$ which are obtained by using the Metropolis-Hastings algorithm for $J\in \{0.0, 0.3, 0.6, 0.9, 1.2, 1.5 \}$ and a $z \in [0.05, 2.00]$ in steps of $\Delta z = 0.05$ where $x_{0,i}$ is sampled from a log-normal distribution with different values for $\sigma$ as indicated in the legend. The solid yellow line corresponds to the line $\langle O \rangle = M\langle L \rangle /2$ and the black curve corresponds to $\langle O \rangle = \langle L \rangle ^2 / N^2 $. Note that in the plots corresponding to $J = 0.0$ and $J = 0.3$ the dark blue and green points are difficult to see because they are stacked upon each other.}
\label{fig:CM sigmas comparison}
\end{figure*}

Figure \ref{fig:CM sigmas comparison} shows the relationship between the average overlap and the number of links in the network with simulation data that was obtained by using the Metropolis-Hastings algorithm for a variety of values for $J$ and $\sigma$. The linear upper limit is illustrated as a solid yellow line. The quadratic lower limit is illustrated as a solid black curve.
This figure shows that in the case where $ J = 0 $ the data points that correspond to $x_{0,i}$ being sampled from a log-normal distribution with a relatively low value for $\sigma$ are either on or close to the quadratic lower limit curve. On the other hand, the case where $\sigma = 10$ (relatively large) results in data points where the overlap in the network for a given number of links is almost maximal and therefore approaches the linear upper limit. When we increase the value of $J$ the data points corresponding to relatively low values of $\sigma$ (e.g. $\sigma = 10^{-5}$ and $\sigma = 10^{-3}$) stay on or close to the quadratic lower limit which is similar to the result in Subsection \ref{subsec: delta distribution} and suggests that the low level of heterogeneity of these networks limits the overlap in the network for a given number of links. The data points corresponding to the intermediate value of $\sigma = 1.0$ however are distributed among a curve similar to the quadratic lower limit curve initially. Increasing the value of $J$ leads to the data points being distributed in a more linear fashion, approaching the linear upper limit. In the case where $\sigma = 10$, the value of $J$ barely influences the value of the overlap for a given number of links since the maximum has almost been reached already when $J = 0.0$. The data therefore shows the effect of increasing $J$ in networks with a high level of homogeneity (or low level of heterogeneity) is a divergence from multiplex configurations with densities of all levels towards multiplex configurations with either low or high density, which is a result of the phase transition as discussed in Subsection \ref{subsec: delta distribution}. It also shows that a very high level of heterogeneity leads to an overlap in the network that is close to maximal for a given number of links. A possible explanation for this was discussed in \ref{subsec: power law distribution}. However, in the case where we have an intermediate level of heterogeneity ($\sigma = 1.0$) we observe that the relationship between the overlap and the number of links in the network transitions from a curve that almost equals the quadratic lower limit curve to the linear upper limit curve when increasing the value of $J$. This means that the effect of the coupling can be relatively strong in the case of a network with an intermediate level of heterogeneity and we can therefore construct networks with a combination of the  overlap and the number of links that falls in between linear upper limit curve and the quadratic lower limit curve in a controlled systematic manner. Note that in Figure \ref{fig:CM sigmas comparison} we can see that when $J$ increases (when $J > 1$) the data points appear to diverge away from values that correspond to intermediate densities. This behavior can also be observed in figure \ref{fig:O L relationship various J} in a more pronounced manner. This is likely due to the fact that as $J$ increases, a larger number of multilinks shall be either in the low density or high density phase.

\begin{figure*}
\begin{center}
\includegraphics[width=.49\textwidth]{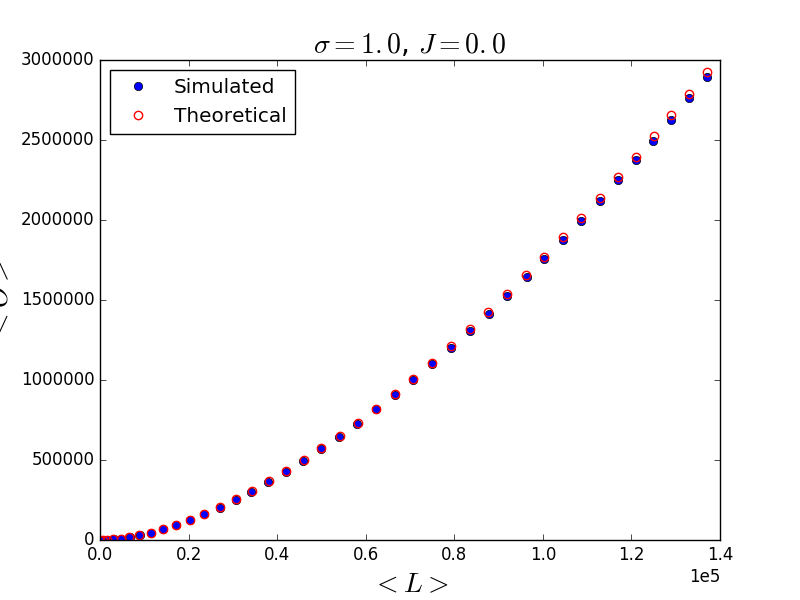}
\includegraphics[width=.49\textwidth]{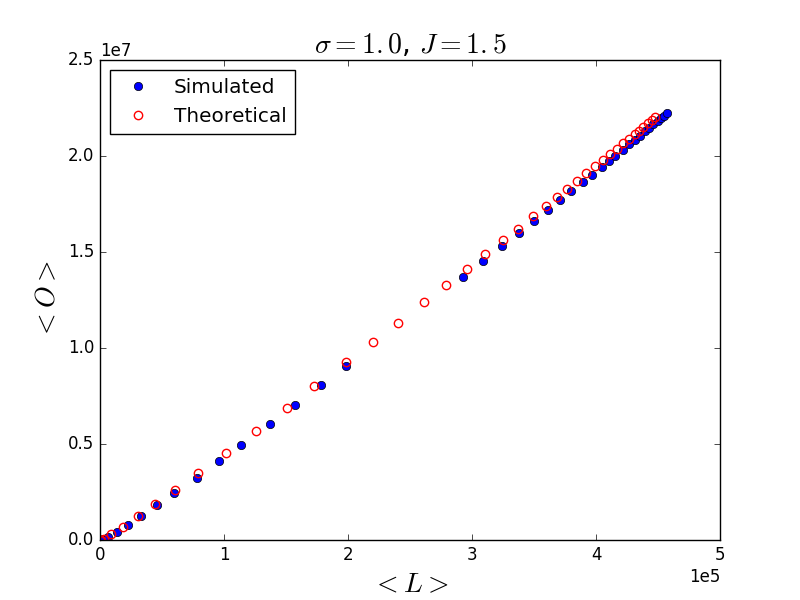}
\end{center}
\caption{The blue points correspond to the mean values of $O$ and $L$ which are obtained by using the Metropolis-Hastings algorithm for $J = 0$ in the left plot and $J = 1.5$ in the right plot and every data point corresponds to a $z \in [0.05, 2.00]$ in steps of $\Delta z = 0.05$ in the case where $x_{i,0}$ is drawn from a log-normal distribution with $\sigma = 1.0$. The red open circles are the theoretically predicted points corresponding to the same $x_i$ and $J$ that are used in the simulations. For these plot's we've chosen that $M = 100$ and $N = 100$.}
\label{fig:O L relationship CM log normal}
\end{figure*}

By using equations \eqref{eq: u_ij tanh equation}, \eqref{eq: MLP equation degrees} and \eqref{eq: MLP equation overlap} we calculate the theoretically predicted values of $\langle O \rangle$ and $\langle L \rangle $ and compare them to the simulation data where $x_{0,i}$ is sampled from a log-normal distribution with $\sigma = 1$. The result is shown in \ref{fig:O L relationship CM log normal}. The results for $\sigma \in \{10^{-5},10^{-3}, 10^{-1}, 10^{1}\}$ are not shown here since relatively low values for $\sigma$ lead to results similar to results obtained in Subsection \ref{subsec: delta distribution} because the distribution from which we sample $x_{0,i}$ approaches a Delta distribution in this case and relatively high values for $\sigma$ lead to results similar to results obtained in Subsection \ref{subsec: power law distribution} because the distribution from which we sample $x_{0,i}$ approaches a power law distribution in this case. Figure \ref{fig:O L relationship CM} shows that the theoretical predictions are in good agreement with the simulation data.

\section{A brief analysis of empirical data}
\label{sec: Empirical Analysis}
In this section we will briefly explore a dataset that represents the multiplex network of international trade. The different layers of this multiplex network represent different commodities. The vertices in this network represent different countries and a link exists between two countries in a given layer if there is trade between them in that commodity. The weight of a link in a given layer signifies the volume of trade in that commodity between two countries. More specifically, the weight of a link in a given layer is the sum of the import and export volume of trade in that commodity measured in thousands of United States dollars. The data includes $N = 206$ countries and $M = 96$ commodities. Some examples of traded commodities are meat, fish, dairy products, coffee, and tobacco. 

\begin{figure*}
\begin{center}
\includegraphics[width=.49\textwidth]{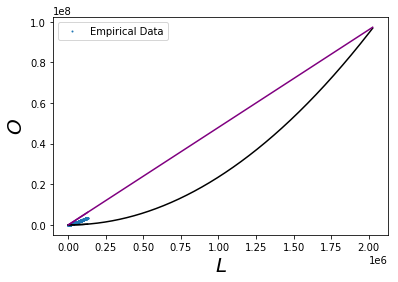}
\includegraphics[width=.49\textwidth]{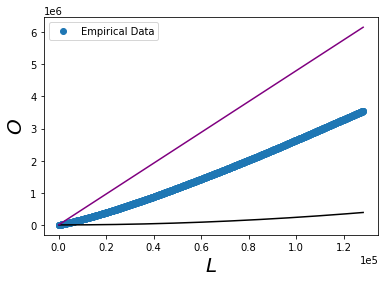}
\end{center}
\caption{The blue points correspond to empirical data points obtained by filtering the multiplex network of international trade. Every data point corresponds to a filtered network which results from choosing a specified number $L^0$ of links in every layer of the international trade network. These $L^0$ links in every layer are the $L^0$ strongest links in the unfiltered trade network. The right figure is a zoomed in version of the left figure. The international trade network consists of $N = 206$ nodes and $M=96$ layers. The purple solid line corresponds to the line $O = ML/2$ and the black curve corresponds to $O = L^2 /N^2$.}
\label{fig:O L relationship empirical data}
\end{figure*}

Using the international trade data, we wish to analyze the overlap by creating $(L,O)$ plots similar to the ones depicted in Figures \ref{fig:O L relationship CM}, \ref{fig:CM powers comparison} or \ref{fig:CM sigmas comparison}. In order to achieve this we repeatedly filter the network such that each layer has the same number of links $L^0 = L^\alpha$ where $\alpha \in \{1,\ldots,M\}$ and calculate the corresponding overlap $O$ for the specified value of $L^0$. The method is to choose the $L^0$ strongest (highest weight) links in every layer. Note that using this filtering method the highest possible density we can achieve is limited by the density of the least dense layer in the unfiltered network. This filtering method was chosen in order to obtain data that is comparable to the model. The result is shown in Figure \ref{fig:O L relationship empirical data}. The figure shows that the overlap for a given number of links appears to be around halfway between the quadratic lower limit curve and the linear upper limit curve. The reason that the overlap for a given number of links is significantly higher than the overlap given by the quadratic lower limit curve is likely due to the heterogeneity of the network, as we have seen that a completely homogeneous network leads to a overlap that follows the quadratic lower limit curve. 

\begin{figure*}
\begin{center}
\includegraphics[width=.7\textwidth]{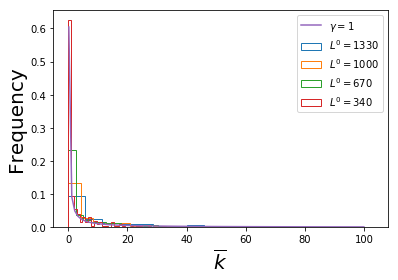}
\end{center}
\caption{The different colors correspond to the histograms of the layer average degrees of filtered networks with different values for $L^0$ (the number of links in each layer). The solid purple curve depicts a power law distribution with $\gamma = 1$.}
\label{fig:empirical degrees}
\end{figure*}

In order to confirm the strong heterogeneity of the network we created a histogram of the layer average degrees in filtered networks with various values for $L^0$ (the number of links in each layer) which is shown in Figure \ref{fig:empirical degrees}. In this figure we included a power law curve with $\gamma = 1$ which  fits the histogram of the degrees for various values of $L^0$, showing that the shape of the degree distribution does not vary with the value of $L^0$. Note however that in order to confirm that the power law curve accurately fits the data, the distributions should be plotted in double logarithmic axes while fitting different possible curves such as the exponential distribution curve. In order to investigate the underlying distribution of the hidden variables $x_i$ we assume for simplicity that $ J = 0$. As we have seen in Equation \eqref{eq: u_ij zero J}, this assumption reduces our model to the Configuration Model. This model was briefly discussed in Section \ref{sec: Example_3}. The maximum likelihood equations in this case are much easier to solve (numerically). When filtering the original network of international trade such that every layer has $L^0$ links, we can find the values of the Configuration Model hidden variables $x_i^*$ of the corresponding filtered network. This procedure is repeated for a range of values for $L^0$.

\begin{figure*}
\begin{center}
\includegraphics[width=.7\textwidth]{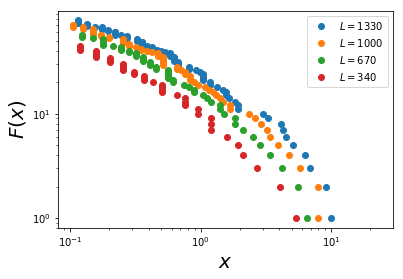}
\end{center}
\caption{This figure shows the cumulative distribution $F(x)$ of the hidden variables $x$ of a filtered network which results from choosing a specified number $L^0$ (as indicated in the legend) of links in every layer of the international trade network. These $L^0$ links in every layer are the $L^0$ strongest links in the unfiltered trade network. $F(x)$ is defined as the number of hidden variables that have a value greater than $x$. }
\label{fig:CM cumulative distribution}
\end{figure*}

The found values of the hidden variables $x_i^*$ can be used in order to plot the cumulative distribution of $x_i^*$ for various values of $L^0$. We choose to plot the cumulative distribution in order to avoid information loss as a result of binning the data. The result is shown in Figure \ref{fig:CM cumulative distribution}. The figure qualitatively shows that the shape of the cumulative distribution of $x$ does not vary with $L^0$. It also shows that the true distribution lies in between a power law distribution (fat tailed) and an exponential distribution, since a power law distribution would appear as a straight line in a double logarithmic plot and an exponential distribution would appear to have a sharp cut off.

\begin{figure*}
\begin{center}
\includegraphics[width=.7\textwidth]{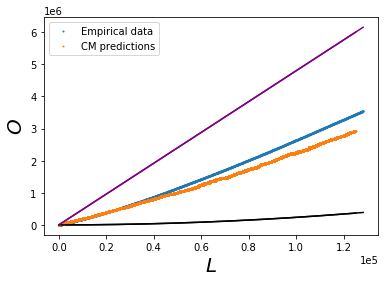}
\end{center}
\caption{The blue points correspond to empirical data points obtained by filtering the multiplex network of international trade. Every data point corresponds to a filtered network which results from choosing a specified number $L^0$ of links in every layer of the international trade network. These $L^0$ links in every layer are the $L^0$ strongest links in the unfiltered trade network. The orange points are the expected number of links and the expected overlap in the network which were obtained by calculating the Configuration Model hidden variables $x_i^*$ corresponding to the different filtered networks. The international trade network consists of $N = 206$ nodes and $M=96$ layers. The purple solid line corresponds to the line $O = ML/2$ and the black curve corresponds to $O = L^2 /N^2$.}
\label{fig:empirical L O CM predictions}
\end{figure*}

\begin{figure*}
\begin{center}
\includegraphics[width=.49\textwidth]{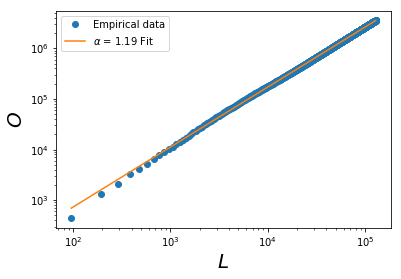}
\includegraphics[width=.49\textwidth]{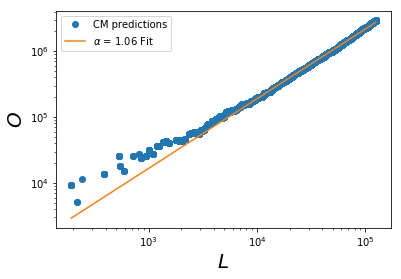}
\end{center}
\caption{The blue points in the left plot correspond to empirical data points obtained by filtering the multiplex network of international trade. Every data point corresponds to a filtered network which results from choosing a specified number $L^0$ of links in every layer of the international trade network. These $L^0$ links in every layer are the $L^0$ strongest links in the unfiltered trade network. The overlap is plotted against the number of links in the network on double logarithmic axes. The yellow line in the left plot corresponds to the curve $O = AL^\alpha$ where $\alpha = 1.19$. The blue points in the right plot are the values of the number of links and the overlap in the network that are predicted by the Configuration Model ($J = 0$) for the different filtered networks. The overlap is plotted against the number of links in the network on double logarithmic axes. The yellow line in the left plot corresponds to the curve $O = AL^\alpha$ where $\alpha = 1.06$.}
\label{fig:O L fits}
\end{figure*}

In Figure \ref{fig:O L relationship empirical data} it can be seen that the filtered networks have a relatively high overlap. The data points appear to be distributed along a similar curve as the simulated data points that correspond to a nonzero $J$ in Figure \ref{fig:CM powers comparison}. We are currently unable to solve the maximum likelihood equations in order to obtain the value of $J$, which shows the necessity to find and/or develop new methods that will allow us to solve the maximum likelihood equations. However, we can use the values of the hidden variables $x_i^*$ corresponding to the data with the assumption that $ J = 0$. By using the values of the hidden variables we can calculate the corresponding expected number of links and the expected overlap in the network which results in a curve. This curve is shown in Figure \ref{fig:empirical L O CM predictions} alongside the curve corresponding to the empirical data. The figure shows that the assumption $J=0$ leads to an insufficiently overlapping network which further demonstrates the necessity of a model that introduces interdependencies between the layers of a network. The difference between the two curves can be quantified by fitting them to the curve 
\begin{equation}
O = AL^\alpha
\end{equation}
where $A$ is a proportionality factor and $\alpha$ is an exponent. We've found the values of $\alpha$ by creating a plot on double logarithmic axes and fitting the lines, which can be seen in Figure \ref{fig:O L fits}. For the empirical data we have found that $\alpha_{\textrm{empirical}} = 1.19$ and for the predictions done by the Configuration Model we have found that $\alpha_{\textrm{CM}} = 1.06$. The difference between the two values is quite small but it is still noticeable. Additionally, this difference may become significant for networks with a higher number of links.

\section{Discussion and conclusions}

In order to better capture the details of a multi-relational system, we propose a minimal model which introduces interdependencies between the layers of a network. In studies of various real world multilayered networks such as the World Trade Network where the  Exponential Random Graph Model (ERGM) was used with the assumption that the layers are independent leads to the observed overlap in the network being significantly different from the overlap predicted by the model \cite{gemmetto2015multiplexity,gemmetto2016multiplexity}. Furthermore, this observed overlap may partly result from spurious correlations and true correlations. In this paper we introduced interdependencies between the layers of a multilayer network in the ERGM by explicitly including the overlap. In order to create a null model of a particular real world network we use the exponential random graph model in combination with the maximum-likelihood method. We have derived the maximum-likelihood equations which theoretically allow us to find the values for the Lagrange multipliers of the ERGM that can be used to generate an exponential random graph ensemble. These values can then be used to create a null model for a particular real world network. 

Given the difficulty of creating a null model by solving the maximum likelihood equations to obtain the values of the Lagrange multipliers corresponding to a particular real network, we treated the Lagrange multipliers as free parameters in order to explore and analyze the properties of multiplex systems as a function of these parameters using numerical methods. Additionally, these numerical results were compared to our analytical results in order to test the validity of the analytical equations that we obtained. We have shown that the analytical equations are very accurate by comparing the theoretically predicted results to the numerically simulated data. In the case of maximally homogeneous networks, the analytical equations predict a global phase transition from a high density phase to a low density phase which was confirmed by the numerical simulations. We have shown that increasing the value of the coupling parameter $J$ generally increases the overlap for a given number of links. However, we have also shown that increasing the heterogeneity of the network increases the overlap for a given number of links as well, which corresponds to increasing the amount of spurious correlations. This is likely a consequence of the presence of large hubs that appear due to  the increased heterogeneity of the network. Additionally, every multilink that is connected to these hubs has a relatively lower critical threshold for the coupling parameter $J$. Therefore, these multilinks have a higher probability to be in the high density phase which leads to a higher overlap as well, which corresponds to increasing the amount of true correlation. The overlap for a given number of links can therefore be increased by either increasing the heterogeneity of the network or the value of the coupling parameter. This can be used in order to create multiplexes with a specific amount of overlap for a given of number of links, given that it is within the theoretical limits discussed in Chapter \ref{sec: Numerical Analysis}. Finally, by using a dataset that represents the multiplex network of international trade we have shown that the assumption that there is no coupling between the layers ($J=0$), which reduces our model to the Configuration Model, results in a insufficiently overlapping network. This means that the empirical overlap is not merely the result of the heterogeneity of the network (spurious correlation between the degrees), which is measured by the Configuration Model, but requires a nonzero coupling (true correlation between the layers). These results demonstrate the necessity of a model that introduces interdependencies between the layers of a network. In this paper we have made a first attempt at proposing such a model. Our model can be seen as a minimal one, to be further generalized in the future.

\appendix 
\section{Hubbard Stratonovich transformation}
\label{A: Pair Partition Function Derivation}
Given the pair Hamiltonian
\begin{equation}
h_{ij} = -B_{ij}\sum_{\alpha=1}^M \sigma_{ij}^\alpha - \frac{J}{2M}\left(\sum_{\alpha=1}^M \sigma_{ij}^\alpha \right)^2 + \frac{J}{2} + h_{ij}^0.
\end{equation}
from Subsection \ref{chapter: 3} we want to obtain an expression for $z_{ij}$ which we defined in Equation \eqref{eq: pair partition function}. By defining $s_{ij} \equiv \{\sigma_{ij}^1,\sigma_{ij}^2,\ldots,\sigma_{ij}^M \}$ as the multilink of $(i,j)$ in terms of the $\sigma$ variables and $\mathcal{S}_{ij}$ as the set containing all $2^{M(M-1)/2}$ possible configurations of $s_{ij}$, the pair partition function can be written as 
\begin{equation}
\begin{split}
z_{ij} & = \sum_{s_{ij}\in \mathcal{S}_{ij}}e^{-h_{ij}} \\
& = \sum_{s_{ij}\in \mathcal{S}} \exp{\Bigg[ \frac{J}{2M}\left(\sum_{\alpha=1}^M \sigma_{ij}^\alpha \right)^2 + B_{ij}\sum_{\alpha=1}^M\sigma_{ij}^\alpha - \frac{J}{2} - h_{ij}^0 \Bigg]} \\
& = e^{-J/2}e^{-h_{ij}^0} \\
& \sum_{s_{ij}\in \mathcal{S}_{ij}} \exp{\left[ \left(\sqrt{\frac{J}{2M}} \sum_{\alpha=1}^M \sigma_{ij}^\alpha \right)^2 + B_{ij}\sum_{\alpha=1}^M\sigma_{ij}^\alpha \right]}.
\end{split}
\end{equation}
The argument of the exponent in the above expression can be linearized by using the Gaussian integral
\begin{equation}
e^{a^2} = \frac{1}{\sqrt{2\pi}}\int^{\infty}_{-\infty}d\xi_{ij} e^{-\xi_{ij}^2/2 + \sqrt{2}a\xi_{ij}}.
\end{equation}
In our case, by choosing $a = \sqrt{J/(2M)}\sum_{\alpha=1}^M\sigma_{ij}^\alpha$ the partition function factorizes with respect to the individual summations over $\sigma_{ij}^\alpha$:
\begin{equation}
\begin{split}
z_{ij}  = & \, \frac{1}{\sqrt{2\pi}}e^{-J/2}e^{-h_{ij}^0}\\
& \sum_{s_{ij}\in \mathcal{S}_{ij}}\int^{\infty}_{-\infty}d\xi_{ij} e^{-\xi_{ij}^2/2} \exp{\left[\sum_{\alpha=1}^M \sigma_{ij}^\alpha \left(\sqrt{\frac{J}{M}}\xi_{ij} + B_{ij} \right) \right]} \\ 
 = & \, \frac{1}{\sqrt{2\pi}}e^{-J/2}e^{-h_{ij}^0} \int^{\infty}_{-\infty} d\xi_{ij} e^{-\xi_{ij}^2/2} 
\\ &
\sum_{\sigma_{ij}^1\in \{-1,1\}}\sum_{\sigma_{ij}^2\in \{-1,1\}} \cdots \sum_{\sigma_{ij}^M\in \{-1,1\}} \\ 
& \prod_{\alpha=1}^M \exp{\left[ \sigma_{ij}^\alpha \left(\sqrt{\frac{J}{M}}\xi_{ij} + B_{ij} \right) \right]} \\ 
 = & \, \frac{2^M}{\sqrt{2\pi}}e^{-J/2}e^{-h_{ij}^0} \\ 
& \int^{\infty}_{-\infty} d\xi_{ij} e^{-\xi_{ij}^2/2} \left[\cosh{\left(\sqrt{\frac{J}{M}}\xi_{ij} + B_{ij} \right)} \right]^M.
\end{split}
\end{equation}
Performing the change of variable $\sqrt{J/M}\xi_{ij} = Jy_{ij}$ we obtain
\begin{equation}
\label{eq: partition function hubbard}
z_{ij} = 2^M \sqrt{\frac{JM}{2\pi}} e^{-J/2}e^{-h_{ij}^0} \int^{\infty}_{-\infty} d\xi_{ij} \left(\Phi_{J,B_{ij}}(y_{ij})\right)^M
\end{equation}
where 
\begin{equation}
\Phi_{J,B_{ij}} \equiv  e^{-Jy_{ij}^2/2}\cosh{\left(Jy_{ij} + B_{ij}\right)}.
\end{equation}
It was previously stated that we are analyzing our system in the large $M$ limit. To proceed in the calculation of $z_{ij}$, it is useful to define the quantity 
\begin{equation}
\label{eq: free energy definition}
f_{ij} \equiv - \lim_{M\rightarrow \infty} \frac{1}{M} \ln{z_{ij}} = - \lim_{M\rightarrow \infty}\ln{z_{ij}^{1/M}}
\end{equation}
which is known as the \textit{free energy} in statistical physics. By inserting the result \eqref{eq: partition function hubbard} into \eqref{eq: free energy definition}, we obtain
\begin{equation}
\begin{split}
f_{ij} = & -\ln{2} - \lim_{M\rightarrow \infty} \frac{1}{M}\ln{\left[ e^{-J/2} \sqrt{\frac{JM}{2\pi}} \right]} + \lim_{M\rightarrow \infty}\frac{h_{ij}^0}{M} \\
& -  \ln{\left[\lim_{M\rightarrow \infty}\left( \int^{\infty}_{-\infty}dy_{ij}\left[\Phi_{J,B_{ij}}(y)\right]^M \right)^{1/M}\right]} \\
 = & -\ln{2} + \frac{J}{2} - B_{ij} \\ 
 & - \ln{\left[\lim_{M\rightarrow \infty}\left( \int^{\infty}_{-\infty}dy_{ij}\left[\Phi_{J,B_{ij}}(y)\right]^M \right)^{1/M}\right]} 
\end{split}
\end{equation}
In order to obtain a more explicit form of the function $f_{ij}$ we use the Laplace theorem \cite{polya1972problems} . Let $\phi(y)$ and $\psi(y)$ be continuous and positive functions within a range $c\leq y \leq d$, then 
\begin{equation}
\lim_{M\rightarrow \infty} \left[ \int^{d}_{c} \psi(y)\left(\phi(y)\right)^M \right]^{1/M} = \max_{c\leq y \leq d}\phi(y).
\end{equation}
For $\psi(y) = 1 $ and $\phi(y) = \Phi_{J,B_{ij}}(y)$ this results in 
\begin{equation}
f_{ij} = -\ln{2} + \frac{J}{2} - B_{ij} - \ln{\left[\max_{-\infty \leq y_{ij} \leq \infty} \Phi_{J, B_{ij}}(y_{ij}) \right]}
\end{equation}
The derivative of $\Phi_{J,B_{ij}}(y_{ij})$ with respect to $y_{ij}$ is zero at its maximum:
\begin{equation}
\begin{split}
\frac{d\Phi_{J,B_{ij}}(y_{ij})}{dy_{ij}} = & \, Je^{-Jy_{ij}^2/2}\sinh{\left(Jy_{ij} + B_{ij} \right)} \\ 
& - Jy_{ij}e^{-Jy_{ij}^2/2}\cosh{\left( Jy_{ij} + B_{ij} \right)} \\
= & \, 0
\end{split}
\end{equation}
The variable $y_{ij}$ therefore obeys the equation
\begin{equation}
\label{eq: self consistent eq y}
y_{ij} = \tanh{\left(Jy_{ij} + B_{ij} \right)}.
\end{equation}
Note that this equation is identical to equation obtained for the magnetization in the Ising Model, and depending on the values of $J$ and $B_{ij}$ there may either be one or three solutions $y_{ij}(J,B_{ij}) \equiv y_{ij}^*$ that satisfy Equation \eqref{eq: self consistent eq y}. The free energy $f_{ij}$ can now be written as a function of $J$ and $B_{ij}$:
\begin{equation}
f_{ij} = -\ln{2} + \frac{J}{2} - B_{ij} + \frac{J}{2}\left(y_{ij}^*\right)^2 - \ln{\left[ \cosh{\left(Jy_{ij}^* + B_{ij} \right)} \right]}.
\end{equation}
We then finally arrive at the pair partition function
\begin{equation}
z_{ij} = e^{-Mf_{ij}} = 2^Me^{-h_{ij}^0}e^{-JM\left(y_{ij}^*\right)^2/2}\cosh^M{\left(Jy_{ij}^* + B_{ij} \right)}
\end{equation}

\FloatBarrier
\bibliography{main}
\end{document}